\documentclass[lettersize,journal]{IEEEtran}
\usepackage{amsmath,amsfonts,amssymb,amsthm}
\usepackage{algorithmic}
\usepackage{algorithm}
\usepackage{array}
\usepackage[caption=false,font=footnotesize,labelfont=rm,textfont=rm]{subfig}
\usepackage{textcomp}
\usepackage{stfloats}
\usepackage{url}
\usepackage{verbatim}
\usepackage{graphicx}
\usepackage{cite}
\usepackage{color}
\begin{document}

\title{Movable Antenna-Aided Near-Field Integrated Sensing and Communication}

\author{Jingze Ding, \IEEEmembership{Graduate Student Member, IEEE},
		Zijian Zhou, \IEEEmembership{Member, IEEE},
		Xiaodan Shao, \IEEEmembership{Member, IEEE},\\
		Bingli Jiao, \IEEEmembership{Senior Member, IEEE},
		and Rui Zhang, \IEEEmembership{Fellow, IEEE}
\thanks{This work was supported in part by the Project of General Administration of Customs of the People's Republic of China under Grant 2024HK290, in part by the Guangzhou Municipal Science and Technology Project under Grants 2023B04J0011 and 2025B04J0038, in part by The Guangdong Provincial Key Laboratory of Big Data Computing, in part by the National Natural Science Foundation of China under Grant 62331022, in part by the 2022 Stable Research Program of Higher Education of China under Grant 20220817144726001, and in part by the Guangdong Major Project of Basic and Applied Basic Research under Grant 2023B0303000001. \textit{ (Corresponding authors: Zijian Zhou; Rui Zhang.)}}
\thanks{Jingze Ding is with the School of Electronics, Peking University, Beijing 100871, China (e-mail: djz@stu.pku.edu.cn).}
\thanks{Zijian Zhou is with the School of Science and Engineering, The Chinese University of Hong Kong, Shenzhen 518172, China (e-mail: zijianzhou@link.cuhk.edu.cn).}
\thanks{Xiaodan Shao is with the Department of Electrical and Computer Engineering, University of Waterloo, Waterloo, ON N2L 3G1, Canada (email: x6shao@uwaterloo.ca).}
\thanks{Bingli Jiao is with the School of Computing and Artificial Intelligence, Fuyao University of Science and Technology, Fuzhou 350109, China. He is also with the School of Electronics, Peking University, Beijing 100871, China (e-mail: jiaobl@pku.edu.cn)}
\thanks{Rui Zhang is with the School of Science and Engineering, Shenzhen Research Institute of Big Data, The Chinese University of Hong Kong, Shenzhen 518172, China (e-mail: rzhang@cuhk.edu.cn). He is also with the Department of Electrical and Computer Engineering, National University of Singapore, Singapore 117583 (e-mail: elezhang@nus.edu.sg).}
}
\maketitle

\begin{abstract}
Integrated sensing and communication (ISAC) is emerging as a pivotal technology for next-generation wireless networks. However, existing ISAC systems are based on fixed-position antennas (FPAs), which inevitably incur a loss in performance when balancing the trade-off between sensing and communication. Movable antenna (MA) technology offers promising potential to enhance ISAC performance by enabling flexible antenna movement. Nevertheless, exploiting more spatial channel variations requires larger antenna moving regions, which may invalidate the conventional far-field assumption for channels between transceivers. Therefore, this paper utilizes the MA to enhance sensing and communication capabilities in near-field ISAC systems, where a full-duplex base station (BS) is equipped with multiple transmit and receive MAs movable in large-size regions to simultaneously sense multiple targets and serve multiple uplink (UL) and downlink (DL) users for communication. We aim to maximize the weighted sum of sensing and communication rates (WSR) by jointly designing the transmit beamformers, sensing signal covariance matrices, receive beamformers, and MA positions at the BS, as well as the UL power allocation. The resulting optimization problem is challenging to solve. Thus, we propose an efficient two-layer random position (RP) algorithm to tackle it. In addition, to reduce movement delay and cost, we design an antenna position matching (APM) algorithm based on the greedy strategy to minimize the total MA movement distance. Extensive simulation results demonstrate the substantial performance improvement achieved by deploying MAs in near-field ISAC systems. Moreover, the results show the effectiveness of the proposed APM algorithm in reducing the antenna movement distance, which is helpful for energy saving and time overhead reduction for MA-aided near-field ISAC systems with large moving regions.
\end{abstract}
\section{Introduction}
\IEEEPARstart{I}{ntegrated} sensing and communication (ISAC) has been considered as a promising technology for next-generation wireless networks because of its unique ability to efficiently reuse time, frequency, power, and hardware resources for both sensing and communication tasks at the same time \cite{mag1,shao}. In addition, the continuous and aggressive utilization of frequency spectrum, such as millimeter-wave (mmWave), in wireless communications has resulted in spectrum overlap with conventional radar systems, thereby driving the need for the development of ISAC frameworks \cite{mag2}.

In the ISAC system, one key challenge is to design dual-functional signals that can achieve both the sensing and communication tasks. It is worth noting that multiple-input multiple-output (MIMO) technology provides a viable solution to this issue by exploiting spatial degrees of freedom (DoFs) through beamforming design. Specifically, MIMO-based ISAC systems, equipped with multiple antennas at both the transmitter and receiver, can employ beamforming to steer the sensing/communication signals toward the desired targets/users, which reduces interference in undesired directions and enhances the quality of ISAC performance \cite{jsac}. Motivated by this, substantial works have explored beamforming design in MIMO-based wireless sensing and communication systems \cite{jsac_onlyone1,jsac_onlyone2,jsac_onlyone3, jsac_onlyone4,jsac_onlyone5,jsac_onlyone6,jsac_onlyone7}. However, most existing ISAC systems focus on either uplink (UL) or downlink (DL) communication, which cannot simultaneously meet both communication demands, thus incurring reduced system throughput. To address this limitation, full-duplex ISAC systems have been proposed \cite{FD}, which enable simultaneous transmission and reception of sensing/communication signals over the same frequency. The full-duplex operation improves both sensing and communication capabilities through the efficient reuse of time-frequency resources. In terms of sensing, the entire frequency bands are available to detect targets so that an enhanced radar performance is achieved. From the communication perspective, there is a significant improvement in spectral efficiency \cite{FD}. Accordingly, full-duplex ISAC systems have garnered significant attention \cite{FD1,FD2,FD3}. The authors in \cite{FD1} investigated the joint secure transceiver design for the full-duplex ISAC system, where the base station (BS) simultaneously performs target tracking and communicates with the UL and DL users. The authors in \cite{FD2} studied the joint optimization of a full-duplex communication-based ISAC system under the criteria of transmit power minimization and sum-rate maximization. The results demonstrated the performance gains in terms of both the power and the spectral efficiency compared to the conventional half-duplex ISAC. More comprehensively, the authors in \cite{FD3} compared the different advantages of ISAC systems operating in full-duplex and half-duplex modes.

However, conventional ISAC architectures mentioned above typically utilize fixed-position antenna (FPA) arrays. The static antenna placement in FPA systems prevents the full exploitation of the wireless channel spatial variation in a given region, due to the lack of local antenna mobility. This limitation hinders the ability to fully optimize spatial diversity and multiplexing performance in sensing and communication tasks, thus constraining the overall potential of ISAC systems \cite{lyu}. Fortunately, movable antenna (MA) and six-dimensional MA (6DMA) technologies have recently been proposed to address this limitation \cite{add_MA1,MA_zhu1,MA_zhu2,add_dingma1,add_6DMA1,6DMA1,6DMA2,6DMA3,6DMA4}. Specifically, MA technology can adjust antenna positions with fixed antenna rotation to effectively provide customized sensing and communication services \cite{add_MA1,MA_zhu1,MA_zhu2,add_dingma1}. More generally, 6DMA technology can incorporate the DoFs in the three-dimensional (3D) position and 3D orientation/rotation of antennas, which can adaptively allocate antenna resources based on the long-term/statistical user distribution to improve network capacity \cite{add_6DMA1,6DMA1,6DMA2,6DMA3,6DMA4}. The various wireless sensing/ISAC systems applying MA have been extensively studied \cite{MA_sens3,MA_sens4,MA_sens5}. The authors in \cite{MA_sens3} analyzed the performance of a new wireless sensing system equipped with a one-dimensional (1D) or two-dimensional (2D) array. The authors in \cite{MA_sens4} minimized the Cram\'er-Rao bound (CRB) through the joint beamforming design and MA position optimization. Moreover, the authors in \cite{MA_sens5} proposed a 6DMA-aided wireless sensing system and compared it with MA for both directive and isotropic antenna radiation patterns. In addition, based on previous studies of MA-aided full-duplex wireless communication systems \cite{MAFD1,MAFD2,MAFD3,MAFD4}, the full-duplex ISAC system aided by MAs has begun to attract the attention of researchers \cite{FD_ISAC1,FD_ISAC2,FD_ISAC3}. The authors in \cite{FD_ISAC1} focused on maximizing the communication rate and sensing mutual information in a monostatic MA-ISAC system. The authors in \cite{FD_ISAC2} investigated the joint discrete antenna positioning and beamforming optimization in MA-enabled full-duplex ISAC networks. The authors in \cite{FD_ISAC3} considered the joint active beamforming and position coefficients design problem in an MA-aided networked full-duplex ISAC system that accomplishes radar sensing as well as UL and DL communication capabilities concurrently. 

While the advantages of MA in ISAC systems have been validated, existing studies mainly focus on far-field ISAC systems. In contrast, the investigation of MA-aided ISAC systems under near-field propagation conditions remains in its infancy \cite{add_near1,add_near2,add_near3}. Generally, to accommodate the free movement of multiple MAs and maximize spatial DoFs, larger antenna moving regions are required \cite{near1,near2}. Hence, the MA system usually has a larger aperture size compared to conventional FPA-based systems. Besides, to meet the ever-growing demands for sensing and communication performance, future ISAC systems are expected to operate in high-frequency bands \cite{near3}. The above two reasons render the conventional far-field assumption commonly adopted in previous MA-aided ISAC systems invalid. As a result, it is essential to explore the potential advantages that MA can offer to ISAC systems in near-field scenarios. It is noteworthy that the additional distance dimension in near-field ISAC compared to far-field ISAC allows the system to provide sensing/communication services for multiple targets/users through joint resolutions in both the angle and distance domains \cite{near4,near5}. However, to the best of the authors' knowledge, there has been no prior work on designing MA-aided full-duplex ISAC systems under near-field channel conditions. Therefore, in this paper, we investigate MA-aided near-field ISAC systems. The main contributions of this paper are summarized as follows:
\begin{itemize}
	\item [1)] 
	We propose an MA-aided ISAC system that employs the near-field spherical wave channel model, where the dual-functional full-duplex BS is equipped with multiple transmit and receive MAs movable in large-size regions to simultaneously sense multiple targets and serve multiple UL and DL users for communication. To balance sensing accuracy and communication efficiency, we aim to maximize the weighted sum of sensing and communication rates (WSR) by jointly designing the transmit beamformers, sensing signal covariance matrices, receive beamformers, and MA positions at the BS, as well as the UL power allocation. 
	\item [2)]
	We propose a two-layer random position (RP) algorithm to solve the formulated non-convex optimization problem with highly-coupled variables. In the inner-layer, for a given MA position, we iteratively update the remaining optimization variables based on the alternating optimization (AO) framework. In the outer-layer, we randomly assign multiple pairs of transmit and receive MA positions and select the pair that maximizes the WSR. Moreover, to reduce the overhead associated with the real-time movement of multiple MAs within the large-size moving region, we propose an antenna position matching (APM) algorithm that effectively minimizes the total MA movement distance.
	\item [3)]
	We conduct extensive simulations to validate the advantages of MA-aided near-field ISAC systems and the effectiveness of the proposed algorithms. The results demonstrate that the MA-aided ISAC system outperforms the ISAC system based on FPAs due to the additional DoF introduced by antenna position optimization. In addition, a larger moving region for MAs increases the equivalent array aperture, thus providing an efficient way to enlarge the near-field region of ISAC systems without increasing the number of antennas, which facilitates multi-location beamfocusing. Furthermore, the proposed APM algorithm effectively reduces the total MA movement distance, which significantly reduces the energy consumption and time overhead for antenna movement in practical systems.
\end{itemize}

The rest of this paper is organized as follows. Section \ref{2} introduces the system model and the optimization problem for the proposed system. In Section \ref{3}, we propose the two-layer RP algorithm and APM algorithm to solve the optimization problem and minimize the total MA movement distance, respectively. Next, simulation results and discussions are provided in Section \ref{4}. Finally, this paper is concluded in Section \ref{5}.

\textit{Notation:}  $a/A$, $\mathbf{a}$, $\mathbf{A}$, and $\mathcal{A}$ denote a scalar, a vector, a matrix, and a set, respectively. $A!$ represents the factorial of positive integer $A$. $\mathbf{A} \succeq \mathbf{0}$ indicates that $\mathbf{A}$ is a positive semidefinite matrix. ${\left(  \cdot  \right)^{T}}$, ${\left(  \cdot  \right)^{H}}$, $\left\|  \cdot  \right\|_2$, $\left|  \cdot  \right|$, $\left\|  \cdot  \right\|_F$, $\mathrm{Tr}\left\lbrace  \cdot\right\rbrace  $, and $\mathrm{Rank}\left\lbrace  \cdot\right\rbrace  $ denote the transpose, conjugate transpose, Euclidean norm, absolute value, Frobenius matrix norm, trace, and rank, respectively. $\mathbb{C}^{M \times N}$ and $\mathbb{R}^{M \times N}$ are the sets for complex and real matrices of $M \times N$ dimensions, respectively. $\mathbf{I}_N$ is the identity matrix of order $N$. $\mathcal{CN}\left( \mathbf{0}, \mathbf{\Lambda} \right) $ represents the circularly symmetric complex Gaussian (CSCG) distribution with mean zero and covariance matrix $\mathbf{\Lambda}$. $\mathcal{A} \backslash \mathcal{B}$ denotes the subtraction of set $\mathcal{B}$ from set $\mathcal{A}$. 
\section{System Model and Problem Formulation}\label{2}
\begin{figure}[!t]
	\centering
	\includegraphics[width=1\linewidth]{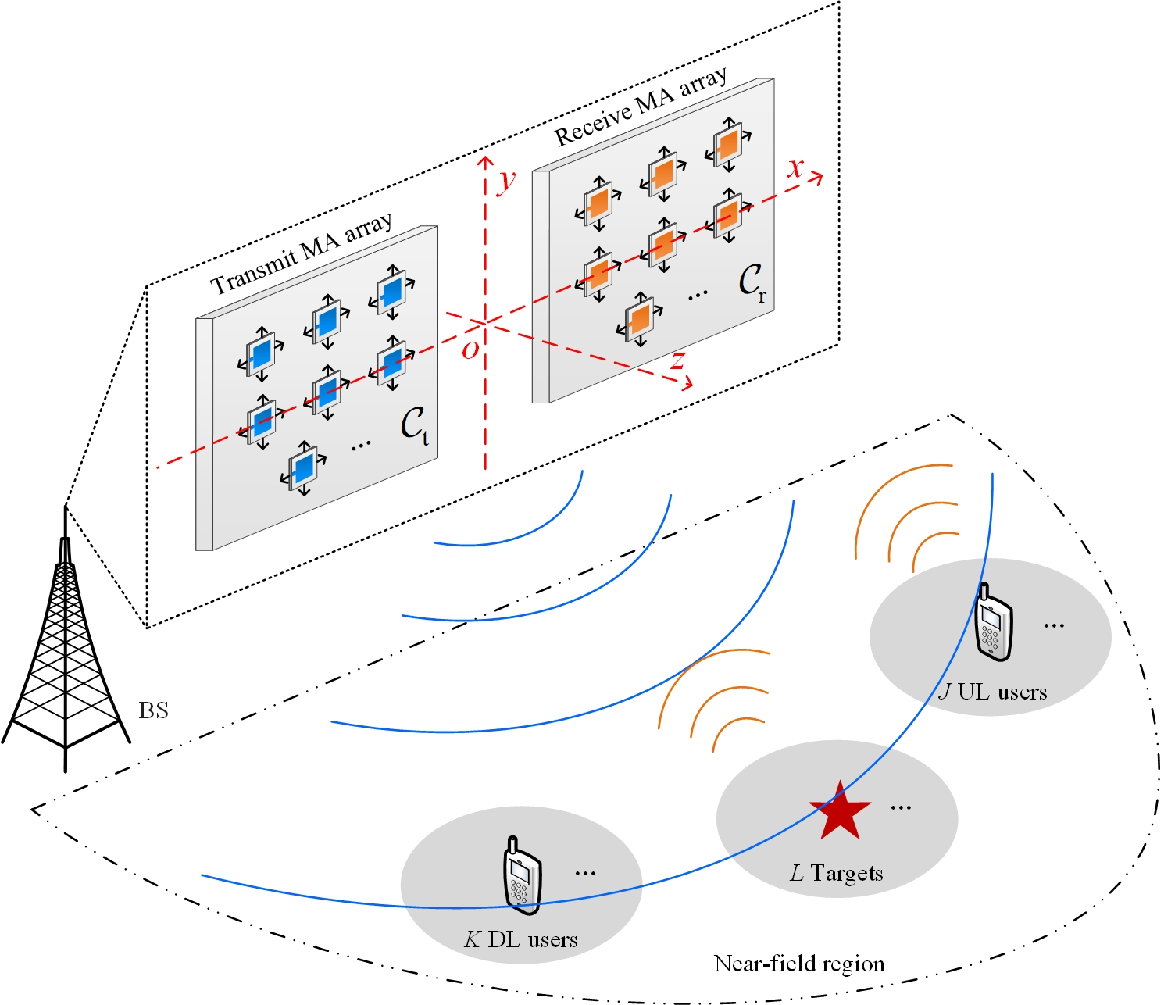}
	\caption{Illustration of the proposed MA-aided near-field ISAC system.}
	\label{system_model}
\end{figure}
Consider an MA-aided near-field ISAC system as shown in Fig. \ref{system_model}, where a dual-functional full-duplex BS equipped with two MA arrays transmits the DL ISAC signal and receives the UL communication signal from $J$ single-FPA half-duplex UL users, along with the reflected ISAC signal from sensing targets, via the same time-frequency resource. The DL ISAC signal is transmitted from the $N$-element MA array to simultaneously communicate with $K$ single-FPA half-duplex DL users and detect $L$ point sensing targets. The sensing echo signal and the UL communication signal are received at the BS through the receive MA array equipped with $M$ elements. Each transmit or receive MA can move freely within its designated transmit or receive region, i.e., $\mathcal{C}_\mathrm{t}$ or $\mathcal{C}_\mathrm{r}$, respectively. Without loss of generality, we assume that $\mathcal{C}_\mathrm{t}$ and $\mathcal{C}_\mathrm{r}$ are square regions with the same side length $A$. Generally, the boundary between near-field and far-field can be determined by the Rayleigh distance \cite{near3}. For the considered MA-aided ISAC system, the Rayleigh distance depends on the size of the moving region and is defined as $4A^2/\lambda$, where $\lambda$ is the carrier wavelength. We assume that the targets and users are located within the BS's near-field region, which implies that their distances from the BS are less than $4A^2/\lambda$.
\subsection{Channel Model}
We consider the quasi-static\footnote{In this paper, the proposed MA-aided near-field ISAC system is designed for scenarios with slow-fading propagation. For fast-fading channels, the MA positions can be optimized using the two-timescale optimization scheme \cite{near2} based on statistical channel state information (CSI) rather than instantaneous CSI to avoid performance degradation caused by CSI variations during MA movement.} near-field spherical wave channel model \cite{near1,near2} for the self-interference (SI) channel, communication channel, and sensing channel\footnote{In typical near-field scenarios, the line-of-sight (LoS) path significantly dominates the non-LoS (NLoS) paths and thus the latter are negligible.}. Since these channels can be actively reconfigured through MA movement, we establish a global Cartesian coordinate system $o$-$xyz$ at the BS to describe the MA positions. The reference point between the transmit and receive regions is defined as origin $o$, where axes $x$ and $y$ are defined as the horizontal and vertical directions in the MA array plane, respectively, and axis $z$ is perpendicular
to the array plane (see Fig. \ref{system_model}). The coordinates of $N$ transmit
MAs and $M$ receive MAs are described by $\mathbf{t} = {\left[ {\mathbf{t}_1^T, \ldots ,\mathbf{t}_N^T} \right]^T} \in {\mathbb{R}^{3N \times 1}}$ and $\mathbf{r} = {\left[ {\mathbf{r}_1^T, \ldots ,\mathbf{r}_M^T} \right]^T} \in {\mathbb{R}^{3M \times 1}}$, where ${\mathbf{t}_n} = \left[ {x_n^\mathrm{t},y_n^\mathrm{t},0} \right]^T\in \mathcal{C}_\mathrm{t}$ ($1 \le n \le N$) and ${\mathbf{r}_m} = \left[ {x_m^\mathrm{r},y_m^\mathrm{r},0} \right]^T \in \mathcal{C}_\mathrm{r}$ ($1 \le m \le M$), respectively. The coordinates of UL user $j$ ($j \in \mathcal{J} = \left\{ {1, \ldots ,J} \right\}$), DL user $k$ ($k \in \mathcal{K} = \left\{ {1, \ldots ,K} \right\}$), and target $l$ ($l \in \mathcal{L} = \left\{ {1, \ldots ,L} \right\}$) are denoted as $\mathbf{q}_{\mathrm{U}_j} \in \mathbb{R}^{3 \times 1}$, $\mathbf{q}_{\mathrm{D}_k} \in \mathbb{R}^{3 \times 1}$, and $\mathbf{q}_{\mathrm{S}_l} \in \mathbb{R}^{3 \times 1}$, respectively. 

In full-duplex systems, the SI signal must be significantly suppressed through effective SI cancellation techniques. Typically, the SI cancellation can be achieved at three stages: passive antenna suppression, active analog cancellation at the radio frequency level, and digital SI cancellation at the baseband level\footnote{Since the analog and digital SI cancellations can actively adapt to channel variations \cite{MAFD3}, the impact of SI channel fluctuations on their performance is neglected in this paper.} \cite{exp_dri}. Therefore, we define $0 < \rho_\mathrm{SI} \ll 1 $ as the SI loss coefficient, which characterizes the overall effectiveness of the three-stage SI cancellation. Based on the uniform-power distance, the channel models suitable for characterizing signal propagation in near-field scenarios can be categorized into the non-uniform spherical wave (NUSW) model and the uniform spherical wave (USW) model \cite{near3}. In the NUSW model, the propagation distance is shorter than the uniform-power distance. As a result, the variations in channel gain become non-negligible. In contrast, when the propagation distance exceeds the uniform-power distance in the USW model, the channel gains are approximately uniform. Without loss of generality, we adopt the USW model. Thus, the SI channel is given by
\begin{align}
	&{\mathbf{H}_\mathrm{SI}}\left( \mathbf{t},\mathbf{r}\right) \nonumber\\
	  &= \rho_\mathrm{SI} \left[ {\begin{array}{*{20}{c}}
			{{e^{\mathrm{j}\frac{{2\pi}}{\lambda }{{\left\| {{\mathbf{t}_1} - {\mathbf{r}_1}} \right\|}_2}}}}& \ldots &{{e^{\mathrm{j}\frac{{2\pi}}{\lambda }{{\left\| {{\mathbf{t}_N} - {\mathbf{r}_1}} \right\|}_2}}}}\\
			\vdots & \vdots & \vdots \\
			{{e^{\mathrm{j}\frac{{2\pi}}{\lambda }{{\left\| {{\mathbf{t}_1} - {\mathbf{r}_M}} \right\|}_2}}}}& \ldots &{{e^{\mathrm{j}\frac{{2\pi}}{\lambda }{{\left\| {{\mathbf{t}_N} - {\mathbf{r}_M}} \right\|}_2}}}}
	\end{array}} \right] \in \mathbb{C}^{M \times N}.
\end{align}
The DL user $k$'s and UL user $j$'s communication channels can be respectively expressed as the functions of the transmit and receive MAs' position vectors, i.e.,
\begin{align}
	& {\mathbf{h}_k}\left(\mathbf{t} \right)  = {\rho _{{\mathrm{D}_k}}}{\left[ {{e^{\mathrm{j}\frac{{2\pi}}{\lambda }{{\left\| {{\mathbf{t}_1} - {\mathbf{q}_{{\mathrm{D}_k}}}} \right\|}_2}}}, \ldots ,{e^{\mathrm{j}\frac{{2\pi}}{\lambda }{{\left\| {{\mathbf{t}_N} - {\mathbf{q}_{{\mathrm{D}_k}}}} \right\|}_2}}}} \right]^T} \in \mathbb{C}^{N \times 1}, \\
	& {\mathbf{f}_j}\left(\mathbf{r} \right) = {\rho _{{\mathrm{U}_j}}}{\left[ {{e^{\mathrm{j}\frac{{2\pi}}{\lambda }{{\left\| {{\mathbf{r}_1} - {\mathbf{q}_{{\mathrm{U}_j}}}} \right\|}_2}}}, \ldots ,{e^{\mathrm{j}\frac{{2\pi}}{\lambda }{{\left\| {{\mathbf{r}_M} - {\mathbf{q}_{{\mathrm{U}_j}}}} \right\|}_2}}}} \right]^T} \in \mathbb{C}^{M \times 1},
\end{align}
where ${\rho _{{\mathrm{D}_k}}}$ and ${\rho _{{\mathrm{U}_j}}}$ represent the corresponding path loss\footnote{To provide a performance upper bound for realistic scenarios and robust designs, this paper assumes that the CSI of ${\mathbf{H}_\mathrm{SI}}$, ${\mathbf{h}_k}$, and ${\mathbf{f}_j}$ is perfectly available at the full-duplex BS.}. 

For the sensing channel of target $l$, we denote the transmit near-field response vector by 
\begin{equation}
	{\mathbf{g}_{\mathrm{t},l}}\left(\mathbf{t} \right) = {\left[ {{e^{\mathrm{j}\frac{{2\pi}}{\lambda }{{\left\| {{\mathbf{t}_1} - {\mathbf{q}_{{\mathrm{S}_l}}}} \right\|}_2}}}, \ldots ,{e^{\mathrm{j}\frac{{2\pi}}{\lambda }{{\left\| {{\mathbf{t}_N} - {\mathbf{q}_{{\mathrm{S}_l}}}} \right\|}_2}}}} \right]^T} \in \mathbb{C}^{N \times 1},
\end{equation}
and similarly denote by
\begin{equation}
	{\mathbf{g}_{\mathrm{r},l}}\left(\mathbf{r} \right) = {\left[ {{e^{\mathrm{j}\frac{{2\pi}}{\lambda }{{\left\| {{\mathbf{r}_1} - {\mathbf{q}_{{\mathrm{S}_l}}}} \right\|}_2}}}, \ldots ,{e^{\mathrm{j}\frac{{2\pi}}{\lambda }{{\left\| {{\mathbf{r}_M} - {\mathbf{q}_{{\mathrm{S}_l}}}} \right\|}_2}}}} \right]^T} \in \mathbb{C}^{M \times 1},
\end{equation}
the receive near-field response vector. The target $l$'s sensing channel is thus given by $\mathbf{G}_l\left( \mathbf{t},\mathbf{r}\right) = {\rho _{{\mathrm{S}_l}}} {\mathbf{g}_{\mathrm{r},l}}\left(\mathbf{r} \right){\mathbf{g}_{\mathrm{t},l}}\left(\mathbf{t} \right)^H \in \mathbb{C}^{M \times N}$, where ${\rho _{{\mathrm{S}_l}}}$ is the round-trip channel coefficient determined by the path loss and the radar cross-section of the target. Following \cite{jsac_onlyone4,jsac_onlyone5,jsac_onlyone6,FD2}, we assume that ${\rho _{{\mathrm{S}_l}}}$ and $\mathbf{q}_{{\mathrm{S}_l}}$ are known or previously estimated at the BS for designing the best suitable sensing waveform to detect this specific target of interest, i.e., target $l$. It is worth noting that the assumption of known ${\rho _{{\mathrm{S}_l}}}$ and $\mathbf{q}_{{\mathrm{S}_l}}$ is idealized and mainly serves to reveal the fundamental performance of MA-aided near-field ISAC systems. In practice, such parameters are usually unavailable a priori and must be estimated via previous observations \cite{add1,addliu}. As a future extension, robust designs that account for the uncertainty of the sensing channel can be developed. Specifically, the CSI uncertainty of sensing channel $\mathbf{G}_l$ can be modeled based on a deterministic model \cite{add2}, i.e., ${\mathbf{G}_l} = {{\widehat {\mathbf{G}}}_l} + \Delta {{ \mathbf{G}}_l}$, where ${{\widehat {\mathbf{G}}}_l}$ is the CSI estimate and $\Delta {{ \mathbf{G}}_l}$ represents the unknown channel uncertainty. The continuous set ${\mathcal{G} _l} = \left\{ {{\mathbf{G}_l} \in {\mathbb{C}^{M \times N}}:{{\left\| {\Delta {\mathbf{G}_l}} \right\|}_F} \le {\varepsilon _l}} \right\}$ contains all possible channel uncertainties with bounded magnitude ${\varepsilon _l}$. The system design with $\Delta {\mathbf{G}_l} \in {\mathcal{G} _l}$ involves infinitely many constraints due to the continuity of the CSI uncertainty set. To enable efficient optimization, these constraints can be equivalently transformed into linear matrix inequality (LMI) constraints using the S-procedure \cite{add3}.
\subsection{Signal Model}
We first focus on the DL ISAC signal used for simultaneous sensing and DL multi-user communication via $N$-element MA array beamforming, which is expressed as  
\begin{equation}
	\mathbf{x} = \sum\limits_{l \in \mathcal{L}} {{\mathbf{s}_l}}+\sum\limits_{k \in \mathcal{K}} {{\mathbf{w}_k}{d_{{\mathrm{D}_k}}}} \in \mathbb{C}^{N \times 1},
\end{equation}
where $\mathbf{w}_k \in \mathbb{C}^{N \times 1}$ is the beamformer of DL user $k$ and ${d_{{\mathrm{D}_k}}}$ is the corresponding DL signal with normalized power, i.e., $\mathbb{E}\left\{ {{{\left| {{d_{{\mathrm{D}_k}}}} \right|}^2}} \right\} = 1$. $\mathbf{s}_l \in \mathbb{C}^{N \times 1}$ is the dedicated sensing signal for target $l$ with the covariance matrix ${\mathbf{S}_l} = \mathbb{E}\left\{ {{\mathbf{s}_l}\mathbf{s}_l^H} \right\} \in \mathbb{C}^{N \times N}$. Here, we assume that the signals $d_{{\mathrm{D}_k}}$ and $\mathbf{s}_l $ are independent of each other. In covariance-based design approaches for MIMO radar systems, the covariance matrix of the waveform is considered instead of the entire waveform \cite{jsac_onlyone5}. Therefore, we focus on the optimization of $\mathbf{S}_l$ in the subsequent analysis. Once $\mathbf{S}_l$ is determined, a dedicated sensing signal $\mathbf{s}_l$ with $\mathbf{S}_l$ as its covariance matrix can be synthesized accordingly \cite{add4}. It should be noted that the synthesized waveform may not satisfy all practical requirements of a real-world radar system, such as the constant-modulus property. The topic of synthesizing practical sensing signals with a given covariance matrix for MA-aided near-field ISAC systems is left for future research. The received signal at DL user $k$ is given by
\begin{equation}\label{y_DL}
	{y_k} = \underbrace {\mathbf{h}_k^H{\mathbf{w}_k}{d_{{\mathrm{D}_k}}}}_\mathrm{Desired\ signal} + \underbrace {\sum\limits_{i \in \mathcal{K}\backslash k} {\mathbf{h}_k^H{\mathbf{w}_i}{d_{{\mathrm{D}_i}}}} }_\mathrm{Multi-user \ interference} + \underbrace {\sum\limits_{l \in \mathcal{L}} {\mathbf{h}_k^H{\mathbf{s}_l}} }_\mathrm{Sensing \ signal} + {n_k},
\end{equation}
where $n_k \sim \mathcal{CN}\left(0,\sigma^2_k \right)  $ represents the additive white Gaussian noise (AWGN) with zero mean and variance $\sigma^2_k$. In \eqref{y_DL}, the co-channel interference from UL to DL users is assumed to be negligible due to the limited transmit power of UL users and the severe attenuation of high-frequency signals that are considered in this paper.

As the full-duplex BS transmits the DL ISAC signal, it simultaneously receives the UL communication signal and the target reflection. Denote the signal from UL user $j$ by ${d_{{\mathrm{U}_j}}}$, which satisfies $\mathbb{E}\left\{ {{{\left| {{d_{{\mathrm{U}_j}}}} \right|}^2}} \right\} = 1$. The received signal at the full-duplex BS can be expressed as
\begin{equation}\label{BS_rece}
	{\mathbf{y}_\mathrm{BS}} = \underbrace {\sum\limits_{j \in \mathcal{J}} {{\mathbf{f}_j}\sqrt {{p_j}} {d_{{\mathrm{U}_j}}}} }_\mathrm{Communication \ signal} + \underbrace {\sum\limits_{l \in \mathcal{L}} {{\mathbf{G}_l}\mathbf{x}} }_\mathrm{Target \ reflection} + \underbrace {{\mathbf{H}_\mathrm{SI}}\mathbf{x}}_\mathrm{SI} + {\mathbf{n}_\mathrm{BS}},
\end{equation}
where $p_j$ is the transmit power of UL user $j$ and ${\mathbf{n}_\mathrm{BS}} \sim \mathcal{CN}\left(\mathbf{0},\sigma^2_\mathrm{BS}\mathbf{I}_M \right)$ is the AWGN at the BS with covariance matrix $\sigma^2_\mathrm{BS}\mathbf{I}_M$.
\subsection{Sensing and Communication Performance Metrics}
The BS uses the received signal \eqref{BS_rece} to sense the target. To capture the reflected signal of target $l$, the BS applies the receive beamformer, $\mathbf{u}_l \in \mathbb{C}^{M \times 1}$, on received signal ${\mathbf{y}_\mathrm{BS}}$, and thus the corresponding signal-to-interference-plus-noise ratio (SINR) is given by
\begin{equation} \label{SINR_Sl}
	{{ \gamma }_{{\mathrm{S}_l}}} = \frac{{\mathbf{u}_l^H{\mathbf{G}_l}\mathbf{R}\mathbf{G}_l^H{\mathbf{u}_l}}}{{\mathbf{u}_l^H\left( {\sum\limits_{j \in \mathcal{J}} {{p_j}{\mathbf{f}_j}\mathbf{f}_j^H}  + {\mathbf{A}_l}\mathbf{R}\mathbf{A}_l^H + \sigma _\mathrm{BS}^2{\mathbf{I}_M}} \right){\mathbf{u}_l}}},
\end{equation}
where $\mathbf{R} = \mathbb{E}\left\{ {\mathbf{x}{\mathbf{x}^H}} \right\} = \sum\nolimits_{l \in \mathcal{L}} {{\mathbf{S}_l}} +\sum\nolimits_{k \in \mathcal{K}} {{\mathbf{w}_k}\mathbf{w}_k^H} \in \mathbb{C}^{N \times N}$ and $\mathbf{A}_l = \sum\nolimits_{i \in \mathcal{L}\backslash l} {{\mathbf{G}_i}}  + {\mathbf{H}_\mathrm{SI}} \in \mathbb{C}^{M \times N}$. For point target detection in MIMO radar systems, an effective radar waveform design strategy is to maximize the output SINR, given that the probability of target detection generally exhibits a monotonic increase with SINR \cite{jsac_onlyone4,FD2}. Although the scalar SINR defined in \eqref{SINR_Sl} inevitably reduces the dimension of observations compared to the full-dimensional data in \eqref{BS_rece}, it provides a tractable and effective surrogate for sensing performance. A sensing system may make measurements of a target to determine its unknown characteristics. To quantify how much information can be extracted with a given sensing signal, the sensing rate, also known as the sensing mutual information per unit time \cite{add5}, can be adopted. It is defined as the mutual information between the received signal at the BS and the target response (or the target parameter), conditioned on the transmitted signal. Specifically, after receive beamforming, the output signal is expressed as $y_{\mathrm{BS}} = \mathbf{u}_l^{H}\mathbf{y}_{\mathrm{BS}}$. Since the BS has knowledge of both the receive beamformer $\mathbf{u}_l$ and the transmitted ISAC signal $\mathbf{x}$, we can use $y_{\mathrm{BS}}$ and \eqref{SINR_Sl} together with the results in \cite{jsac_onlyone7,lyu} to obtain the conditional sensing mutual information between the received signal and the target response (or the sensing rate) as 
\begin{equation}
	R_{\mathrm{S}_l}= I\left(y_{\mathrm{BS}};\mathbf{G}_l \mid \mathbf{u}_l,\mathbf{x}\right) = \log_{2}\left( 1+\gamma_{\mathrm{S}_l}\right).
\end{equation}
The target response $\mathbf{G}_l$ contains the geometric parameters of the target, such as distance and angle information. It has been proved in \cite{addd1} that a high sensing rate indicates that the sensing system can accurately estimate the geometric parameters of targets embedded in the target response matrix. Therefore, we can maximize the sensing rate to obtain more information about the measured target and reduce the measurement error, which finally results in more accurate target parameter estimation \cite{addd1}. Different from the communication rate, the sensing rate has no explicit operational meaning. However, it has similar mathematical properties and the same unit of measurement as the communication rate, which facilitates theoretical analysis and waveform design. The sensing rate has been widely recognized as an effective metric in previous studies \cite{jsac_onlyone7,lyu,FD_ISAC1,near5}.

Similarly, the BS applies another set of receive beamformers, $\mathbf{b}_j \in \mathbb{C}^{M \times 1}$, on ${\mathbf{y}_\mathrm{BS}}$ to decode the data signal of UL user $j$. The corresponding UL communication rate is given by $R_{{\mathrm{U}_j}}=\log_2\left(1+{\gamma _{{\mathrm{U}_j}}} \right) $, where ${\gamma _{{\mathrm{U}_j}}}$ is the receive SINR given by
\begin{equation} \label{SINR_Uj}
	{{ \gamma }_{{\mathrm{U}_j}}} = \frac{{{p_j}\mathbf{b}_j^H{\mathbf{f}_j}\mathbf{f}_j^H{\mathbf{b}_j}}}{{\mathbf{b}_j^H\left( {\sum\limits_{i \in \mathcal{J}\backslash j} {{p_i}{\mathbf{f}_i}\mathbf{f}_i^H}  + \mathbf{A}\mathbf{R}{\mathbf{A}^H} + \sigma _\mathrm{BS}^2{\mathbf{I}_M}} \right){\mathbf{b}_j}}},
\end{equation}
where $\mathbf{A} = \sum\nolimits_{l \in \mathcal{L}} {{\mathbf{G}_l}}  + {\mathbf{H}_\mathrm{SI}} \in \mathbb{C}^{M \times N}$. For DL communication, \eqref{y_DL} indicates that the SINR of DL user $k$ can be expressed as
\begin{equation}\label{SINR_Dk}
	{\gamma _{{\mathrm{D}_k}}} = \frac{{{{\left| {\mathbf{h}_k^H{\mathbf{w}_k}} \right|}^2}}}{{\sum\limits_{i \in \mathcal{K}\backslash k} {{{\left| {\mathbf{h}_k^H{\mathbf{w}_i}} \right|}^2}}  + \sum\limits_{l \in \mathcal{L}} {\mathbf{h}_k^H{\mathbf{S}_l}{\mathbf{h}_k}}  + \sigma _k^2}},
\end{equation}
and the corresponding DL communication rate is given by $R_{{\mathrm{D}_k}}=\log_2\left(1+{\gamma _{{\mathrm{D}_k}}} \right) $.
\subsection{Problem Formulation}
In this paper, we aim to maximize the WSR to balance sensing accuracy and communication efficiency, which can be expressed as
\begin{equation}\label{sum_rate}
	WSR = \sum\limits_{l \in \mathcal{L}} {{\varpi _{{\mathrm{S}_l}}}{R_{{\mathrm{S}_l}}}}  + \sum\limits_{j \in \mathcal{J}} {{\varpi _{{\mathrm{U}_j}}}{R_{{\mathrm{U}_j}}}}  + \sum\limits_{k \in \mathcal{K}} {{\varpi _{{\mathrm{D}_k}}}{R_{{\mathrm{D}_k}}}},
\end{equation}
where ${\varpi _{{\mathrm{S}_l}}} \ge0$, ${\varpi _{{\mathrm{U}_j}}}\ge0$, and ${\varpi _{{\mathrm{D}_k}}}\ge0$ denote predefined rate weights for target $l$, UL user $j$, and DL user $k$, respectively, which satisfy $\sum\nolimits_{l \in \mathcal{L}} {{\varpi _{{\mathrm{S}_l}}}}  + \sum\nolimits_{j \in \mathcal{J}} {{\varpi _{{\mathrm{U}_j}}}}  + \sum\nolimits_{k \in \mathcal{K}} {{\varpi _{{\mathrm{D}_k}}}}=1 $ and can be used to prioritize the targets and users. In particular, we jointly optimize the receive beamformers, $\mathbf{u}_l$ and $\mathbf{b}_j$, sensing signal covariance matrices, $\mathbf{S}_l$, transmit beamformers, $\mathbf{w}_k$, UL transmit power, $p_j$, and MA positions, $\mathbf{t}$ and $\mathbf{r}$. Accordingly, the optimization problem is formulated as\footnote{This paper focuses on maximizing the overall WSR of the proposed MA-aided near-field ISAC system. The quality-of-service (QoS) requirements of individual users and targets can be addressed in subsequent stages via user scheduling and resource allocation.}
\begin{align} \label{max1}
	& \mathop {\mathrm{maximize} }\limits_{ \mathbf{u}_l,\mathbf{b}_j,\mathbf{S}_l,\mathbf{w}_k,p_j,\mathbf{t},\mathbf{r}} \quad WSR \\
	&\mathrm{s.t.} \quad \mathrm{C1}: \left\| {\mathbf{u}_l} \right\|_2^2 = 1 , \ \forall {l} \in \mathcal{L}, \nonumber\\
	&\hspace{2.3em} \mathrm{C2}: \left\| {\mathbf{b}_j} \right\|_2^2 = 1 , \ \forall {j} \in \mathcal{J},\nonumber\\
	&\hspace{2.3em} \mathrm{C3}: \sum\limits_{l \in {\mathcal{L}}} {\mathrm{Tr}\left\{ \mathbf{S}_l \right\}}  + \sum\limits_{k \in {\mathcal{K}}} {\left\| {\mathbf{w}_k} \right\|_2^2} \le P_\mathrm{D}^\mathrm{max}, \nonumber\\
	&\hspace{2.3em} \mathrm{C4}: 0 \le {p_j} \le P_\mathrm{U}^\mathrm{max} , \ \forall j \in \mathcal{J}, \nonumber\\
	&\hspace{2.3em} \mathrm{C5}: \mathbf{t} \in {\mathcal{C}_\mathrm{t}}, \ \mathbf{r} \in {\mathcal{C}_\mathrm{r}}, \nonumber\\
	&\hspace{2.3em} \mathrm{C6}: {\left\| {{\mathbf{t}_a} - {\mathbf{t}_{\tilde a}}} \right\|_2} \ge D , \ 1 \le a \ne {\tilde a} \le N, \nonumber\\
	&\hspace{2.3em} \mathrm{C7}: {\left\| {{\mathbf{r}_b} - {\mathbf{r}_{\tilde b}}} \right\|_2} \ge D,\ 1 \le b \ne {\tilde b} \le M. \nonumber
\end{align}
Here, constraints C1 and C2 normalize the receive beamformers. Constraints C3 and C4 indicate that the total transmit powers of DL and UL transmissions should not exceed the maximum limits, $P_\mathrm{D}^\mathrm{max}$ and $P_\mathrm{U}^\mathrm{max}$, respectively. Constraint C5 confines the moving regions of transmit and receive MAs. Constraints C6 and C7 ensure the minimum inter-MA distance, $D$, at the BS for practical implementation. Note that problem \eqref{max1} is a non-convex optimization problem with coupled variables, and thus finding globally optimal solutions for it in polynomial time is challenging. Thus, we develop an efficient two-layer RP algorithm to obtain suboptimal solutions for this problem in the next section. 
\section{Proposed Solution}\label{3}
The AO algorithm is commonly used to solve optimization problems in wireless communication systems. It decomposes the original problem into manageable sub-problems and iteratively solves each one while keeping the optimization variables of other sub-problems fixed. For MA-aided communication systems, a straightforward approach is to separate the optimization of MA positions and other variables into two independent problems and then solve them iteratively \cite{MA_zhu1}. However, the conventional AO algorithm may converge to undesired local optimal solutions because the MA positions (or other variables) determined in the previous iteration restrict the optimization space for other variables (or MA positions) in the current iteration \cite{hu_AO}. Therefore, we propose a two-layer RP algorithm. In the inner-layer, for a given MA position, we decompose problem \eqref{max1} into two sub-problems, i.e., iteratively updating $\left\{\mathbf{u}_l,\mathbf{b}_j  \right\}$ with closed-form expressions and $\left\{\mathbf{S}_l,\mathbf{w}_k,p_j \right\}$ based on successive convex approximation (SCA). In the outer-layer, we randomly assign multiple pairs of transmit and receive MA positions, $\left\{\mathbf{t},\mathbf{r} \right\}$, and select the pair that maximizes the objective value \eqref{sum_rate} as the optimized MA positions. The initial and optimized MA positions are then matched one by one via the proposed APM algorithm to minimize the total MA movement distance. The details of the proposed algorithms are presented below.
\subsection{Inner-Layer of RP Algorithm}
In the inner-layer, since the MA positions, $\left\{\mathbf{t},\mathbf{r} \right\}$, are given, we only need to optimize $\left\{\mathbf{u}_l,\mathbf{b}_j,\mathbf{S}_l,\mathbf{w}_k,p_j \right\}$. Thus, problem \eqref{max1} can be restated as the following optimization problem: 
\begin{align} \label{max2}
	& \mathop {\mathrm{maximize} }\limits_{ \mathbf{u}_l,\mathbf{b}_j,\mathbf{S}_l,\mathbf{w}_k,p_j} \quad WSR \\
	&\mathrm{s.t.} \quad \mathrm{C1}-\mathrm{C4}. \nonumber
\end{align}
Based on the AO framework, we decompose problem \eqref{max2} into two sub-problems and iteratively optimize $\left\{\mathbf{u}_l,\mathbf{b}_j  \right\}$ and $\left\{\mathbf{S}_l,\mathbf{w}_k,p_j \right\}$.
\subsubsection{Sub-problem 1 for Optimizing $\left\{\mathbf{u}_l,\mathbf{b}_j  \right\}$}
Given $\left\{\mathbf{S}_l,\mathbf{w}_k,p_j \right\}$, the optimizations of $\mathbf{u}_l$ and $\mathbf{b}_j$ only affect the receive SINRs \eqref{SINR_Sl} and \eqref{SINR_Uj}, respectively. Therefore, maximizing the WSR, i.e., objective value \eqref{sum_rate}, is equivalent to maximizing SINRs \eqref{SINR_Sl} and \eqref{SINR_Uj}. Hence, we optimize $\left\{\mathbf{u}_l,\mathbf{b}_j  \right\}$ via the SINR maximization criteria: 
\begin{align}  \label{max_ul}
	& \mathop {\mathrm{maximize} }\limits_{\mathbf{u}_l} \quad {{ \gamma }_{{\mathrm{S}_l}}} \\
	& \mathrm{s.t.} \quad \mathrm{C1}, \nonumber
\end{align}
\begin{align} \label{max_bj}
	& \mathop {\mathrm{maximize} }\limits_{\mathbf{b}_j} \quad {{ \gamma }_{{\mathrm{U}_j}}} \\
	& \mathrm{s.t.} \quad \mathrm{C2}. \nonumber
\end{align}
\newtheorem{theorem}{Proposition}
\begin{theorem}
The optimal solutions of problems \eqref{max_ul} and \eqref{max_bj} are respectively given by
\begin{align}
	&\mathbf{u}_l^* = \frac{{{{\left( {\sum\limits_{j \in \mathcal{J}} {{{\tilde {\mathbf{f}}}_j}\tilde {\mathbf{f}}_j^H}  + {\mathbf{A}_l}\mathbf{R}\mathbf{A}_l^H + \sigma _\mathrm{BS}^2{\mathbf{I}_M}} \right)}^{ - 1}}{\mathbf{g}_l}}}{{{{\left\| {{{\left( {\sum\limits_{j \in \mathcal{J}} {{{\tilde {\mathbf{f}}}_j}\tilde {\mathbf{f}}_j^H}  + {\mathbf{A}_l}\mathbf{R}\mathbf{A}_l^H + \sigma _\mathrm{BS}^2{\mathbf{I}_M}} \right)}^{ - 1}}{\mathbf{g}_l}} \right\|}_2}}}, \label{receive_beam1}\\
	&\mathbf{b}_j^* = \frac{{{{\left( {\sum\limits_{i \in \mathcal{J}\backslash j} {{{\tilde {\mathbf{f}}}_i}\tilde {\mathbf{f}}_i^H}  + \mathbf{A}\mathbf{R}{\mathbf{A}^H} + \sigma _\mathrm{BS}^2{\mathbf{I}_M}} \right)}^{ - 1}}{{\tilde {\mathbf{f}}}_j}}}{{{{\left\| {{{\left( {\sum\limits_{i \in \mathcal{J}\backslash j} {{{\tilde {\mathbf{f}}}_i}\tilde {\mathbf{f}}_i^H}  + \mathbf{A}\mathbf{R}{\mathbf{A}^H} + \sigma _\mathrm{BS}^2{\mathbf{I}_M}} \right)}^{ - 1}}{{\tilde {\mathbf{f}}}_j}} \right\|}_2}}}, \label{receive_beam2}
\end{align}
where ${\mathbf{g}_l} = {\mathbf{G}_l}\left( {\sum\nolimits_{k \in \mathcal{K}} {{\mathbf{w}_k}}  + \sum\nolimits_{l \in \mathcal{L}} {{\mathbf{s}_l}} } \right) \in \mathbb{C}^{M \times 1}$ and ${{\tilde {\mathbf{f}}}_j} = \sqrt {{p_j}} {{\mathbf{f}}_j} \in \mathbb{C}^{M \times 1}$.
\end{theorem}
\begin{IEEEproof}
	Please refer to \cite[Appendix A]{FD2}.
\end{IEEEproof}
\subsubsection{Sub-problem 2 for Optimizing $\left\{\mathbf{S}_l,\mathbf{w}_k,p_j \right\}$}
Given $\left\{\mathbf{u}_l,\mathbf{b}_j  \right\}$, the joint optimization of $\left\{\mathbf{S}_l,\mathbf{w}_k,p_j \right\}$ can be formulated as
\begin{align} \label{max_sca1}
	& \mathop {\mathrm{maximize} }\limits_{\mathbf{S}_l,\mathbf{w}_k,p_j} \quad WSR \\
	&\mathrm{s.t.} \quad \mathrm{C3},\mathrm{C4}. \nonumber
\end{align}
Defining $\mathbf{W}_k = \mathbf{w}_k\mathbf{w}_k^H \in \mathbb{C}^{N \times N}$, constraint C3 can be equivalently transformed into the following constraints:
\begin{align}
	&\mathrm{C3a}: \sum\limits_{l \in {\mathcal{L}}} {\mathrm{Tr}\left\{ \mathbf{S}_l \right\}}  + \sum\limits_{k \in {\mathcal{K}}} {\mathrm{Tr}\left\{ \mathbf{W}_k \right\}} \le P_\mathrm{D}^\mathrm{max}, \nonumber\\
	&\mathrm{C3b}: \mathbf{W}_k \succeq \mathbf{0}, \ \forall {k} \in \mathcal{K}, \nonumber\\
	&\mathrm{C3c}: \mathrm{Rank}\left\{\mathbf{W}_k \right\} \le 1, \ \forall {k} \in \mathcal{K}. \nonumber
\end{align}
As such, problem \eqref{max_sca1} can be recast as
\begin{align} \label{max_sca2}
	& \mathop {\mathrm{maximize} }\limits_{\mathbf{S}_l,\mathbf{W}_k,p_j} \quad WSR \\
	&\mathrm{s.t.} \quad \mathrm{C3a},\mathrm{C3b},\mathrm{C3c},\mathrm{C4}. \nonumber
\end{align}
Problem \eqref{max_sca2} is non-convex due to the non-concavity of the objective function and the rank constraint C3c. Therefore, it is necessary to transform problem \eqref{max_sca2} into a convex form. 

To achieve this goal, we begin by addressing the non-concavity of the objective function. Based on the rule of the logarithmic function, the objective function of problem \eqref{max_sca2} can be reformulated as
\begin{equation}\label{ccv_R}
	\widetilde{WSR} = {\alpha _1} + {\alpha _2} + {\alpha _3} - {\beta _1} - {\beta _2} - {\beta _3},
\end{equation}
where ${\alpha _1}$, ${\alpha _2}$, ${\alpha _3}$, ${\beta _1}$, ${\beta _2}$, and ${\beta _3}$ are the concave functions with respect to (w.r.t.) optimization variables $\left\{\mathbf{S}_l,\mathbf{W}_k,p_j \right\}$ and shown in \eqref{alpha_1}-\eqref{beta_3} at the top of the next page, respectively.
\begin{figure*}[!t]
	\textsc{\centering
		\begin{align}
			& \alpha_1 = \sum\limits_{l \in \mathcal{L}} {{\varpi _{{\mathrm{S}_l}}}{{\log }_2}\left( {\sum\limits_{i \in \mathcal{L}} {\mathrm{Tr}\left\{ {{\mathbf{R}}\mathbf{G}_i^H{\mathbf{u}_l}\mathbf{u}_l^H{\mathbf{G}_i}} \right\} + \mathrm{Tr}\left\{ {{\mathbf{R}}\mathbf{H}_\mathrm{SI}^H{\mathbf{u}_l}\mathbf{u}_l^H{\mathbf{H}_\mathrm{SI}}} \right\} + \sum\limits_{j \in \mathcal{J}} {{p_j}{{\left| {\mathbf{u}_l^H{\mathbf{f}_j}} \right|}^2}}  + \sigma _\mathrm{BS}^2} } \right)} . \label{alpha_1} \\
			& {\alpha _2} = \sum\limits_{j \in \mathcal{J}} {{\varpi _{{\mathrm{U}_j}}}{{\log }_2}\left( {\sum\limits_{l \in \mathcal{L}} {\mathrm{Tr}\left\{ {{\mathbf{R}}\mathbf{G}_l^H{\mathbf{b}_j}\mathbf{b}_j^H{\mathbf{G}_l}} \right\}}  + \mathrm{Tr}\left\{ {{\mathbf{R}}\mathbf{H}_\mathrm{SI}^H{\mathbf{b}_j}\mathbf{b}_j^H{\mathbf{H}_\mathrm{SI}}} \right\} + \sum\limits_{i \in \mathcal{J}} {{p_i}{{\left| {\mathbf{b}_j^H{\mathbf{f}_i}} \right|}^2}}  + \sigma _\mathrm{BS}^2} \right)}. \\ 
			& {\alpha _3} = \sum\limits_{k \in \mathcal{K}} {{\varpi _{{\mathrm{D}_k}}}{{\log }_2}\left( {\mathrm{Tr}\left\{ {{\mathbf{R}}{\mathbf{h}_k}\mathbf{h}_k^H} \right\} + \sigma _k^2} \right)} .\\
			& {\beta _1} = \sum\limits_{l \in \mathcal{L}} {{\varpi _{{\mathrm{S}_l}}}{{\log }_2}\left( {\sum\limits_{i \in \mathcal{L}\backslash l} {\mathrm{Tr}\left\{ {{\mathbf{R}}\mathbf{G}_i^H{\mathbf{u}_l}\mathbf{u}_l^H{\mathbf{G}_i}} \right\} + \mathrm{Tr}\left\{ {{\mathbf{R}}\mathbf{H}_\mathrm{SI}^H{\mathbf{u}_l}\mathbf{u}_l^H{\mathbf{H}_\mathrm{SI}}} \right\} + \sum\limits_{j \in \mathcal{J}} {{p_j}{{\left| {\mathbf{u}_l^H{\mathbf{f}_j}} \right|}^2}}  + \sigma _\mathrm{BS}^2} } \right)} .\\
			& {\beta _2} = \sum\limits_{j \in \mathcal{J}} {{\varpi _{{\mathrm{U}_j}}}{{\log }_2}\left( {\sum\limits_{l \in \mathcal{L}} {\mathrm{Tr}\left\{ {{\mathbf{R}}\mathbf{G}_l^H{\mathbf{b}_j}\mathbf{b}_j^H{\mathbf{G}_l}} \right\}}  + \mathrm{Tr}\left\{ {{\mathbf{R}}\mathbf{H}_\mathrm{SI}^H{\mathbf{b}_j}\mathbf{b}_j^H{\mathbf{H}_\mathrm{SI}}} \right\} + \sum\limits_{i \in \mathcal{J} \backslash j} {{p_i}{{\left| {\mathbf{b}_j^H{\mathbf{f}_i}} \right|}^2}}  + \sigma _\mathrm{BS}^2} \right)}. \\ 
			& {\beta _3} = \sum\limits_{k \in \mathcal{K}} {{\varpi _{{\mathrm{D}_k}}}{{\log }_2}\left( {\mathrm{Tr}\left( {\left( {\sum\limits_{l \in \mathcal{L}} {{\mathbf{S}_l}}  + \sum\limits_{i \in \mathcal{K} \backslash k} {{\mathbf{W}_i}} } \right){\mathbf{h}_k}\mathbf{h}_k^H} \right) + \sigma _k^2} \right)} . \label{beta_3}
	\end{align}\hrulefill}
\end{figure*}
Thus, objective function \eqref{ccv_R} is a difference-of-concave function. The SCA algorithm is applied to obtain a suboptimal solution for problem \eqref{max_sca2}.

Define the maximum number of iterations for SCA as $C$. In the $c$-th ($1 \le c \le C$) iteration, given the feasible point $\left\{\mathbf{S}_l^c ,\mathbf{W}_k^c,p_j^c \right\}$, we construct a global overestimate of ${\beta _1}\left( {{\mathbf{S}_l},{\mathbf{W}_k},{p_j}} \right)$ by the first-order Taylor expansion, i.e., 
\begin{align}
	& {\beta _1}\left( {{\mathbf{S}_l},{\mathbf{W}_k},{p_j}} \right) \le {{\hat \beta }_1}\left( {{\mathbf{S}_l},{\mathbf{W}_k},{p_j}\left| {\mathbf{S}_l^c,\mathbf{W}_k^c,p_j^c} \right.} \right) \nonumber\\
	& = {\beta _1}\left( {\mathbf{S}_l^c,\mathbf{W}_k^c,p_j^c} \right) \nonumber\\
	& + \sum\limits_{l \in \mathcal{L}} {\mathrm{Tr}\left\{ {{{\left( {{\nabla _{{\mathbf{S}_l}}}{\beta _1}\left( {\mathbf{S}_l^c,\mathbf{W}_k^c,p_j^c} \right)} \right)}^H}\left( {{\mathbf{S}_l} - \mathbf{S}_l^c} \right)} \right\}} \nonumber\\  
	& + \sum\limits_{k \in \mathcal{K}} {\mathrm{Tr}\left\{ {{{\left( {{\nabla _{{\mathbf{W}_k}}}{\beta _1}\left( {\mathbf{S}_l^c,\mathbf{W}_k^c,p_j^c} \right)} \right)}^H}\left( {{\mathbf{W}_k} - \mathbf{W}_k^c} \right)} \right\}}  \nonumber\\ 
	& + \sum\limits_{j \in \mathcal{J}} {{\nabla _{{p_j}}}{\beta _1}\left( {\mathbf{S}_l^c,\mathbf{W}_k^c,p_j^c} \right)\left( {{p_j} - p_j^c} \right)} ,
\end{align}
where ${\nabla _{{\mathbf{S}_l}}}{\beta _1}$, ${\nabla _{{\mathbf{W}_k}}}{\beta _1}$, and ${\nabla _{p_j}}{\beta _1}$ are the gradients of function ${\beta _1}$ w.r.t. ${{\mathbf{S}_l}}$, ${{\mathbf{W}_k}}$, and ${p_j}$, respectively, and are shown in \eqref{grad_11} and \eqref{grad_12} at the top of this page.
\begin{figure*}[!t]
	\textsc{\centering
		\begin{align}
			&{\nabla _{{\mathbf{S}_l}}}{\beta _1}={\nabla _{{\mathbf{W}_k}}}{\beta _1} = \sum\limits_{l \in \mathcal{L}} {\frac{{{\varpi _{{\mathrm{S}_l}}}}}{{\ln 2}} \cdot \frac{{\sum\limits_{i \in \mathcal{L}\backslash l} {\mathbf{G}_i^H{\mathbf{u}_l}\mathbf{u}_l^H{\mathbf{G}_i}}  + \mathbf{H}_\mathrm{SI}^H{\mathbf{u}_l}\mathbf{u}_l^H{\mathbf{H}_\mathrm{SI}}}}{{\sum\limits_{i \in \mathcal{L}\backslash l} {\mathrm{Tr}\left\{ { {\mathbf{R}}\mathbf{G}_i^H{\mathbf{u}_l}\mathbf{u}_l^H{\mathbf{G}_i}} \right\} + \mathrm{Tr}\left\{ { {\mathbf{R}}\mathbf{H}_\mathrm{SI}^H{\mathbf{u}_l}\mathbf{u}_l^H{\mathbf{H}_\mathrm{SI}}} \right\} +\sum\limits_{j \in \mathcal{J}} {{p_j}{{\left| {\mathbf{u}_l^H{\mathbf{f}_j}} \right|}^2}}  + \sigma _\mathrm{BS}^2} }}} .\label{grad_11}\\
			&{\nabla _{p_j}}{\beta _1}= \sum\limits_{l \in \mathcal{L}} {\frac{{{\varpi _{{\mathrm{S}_l}}}}}{{\ln 2}} \cdot \frac{{{\left| {\mathbf{u}_l^H{\mathbf{f}_j}} \right|}^2}}{{\sum\limits_{i \in \mathcal{L}\backslash l} {\mathrm{Tr}\left\{ { {\mathbf{R}}\mathbf{G}_i^H{\mathbf{u}_l}\mathbf{u}_l^H{\mathbf{G}_i}} \right\} + \mathrm{Tr}\left\{ { {\mathbf{R}}\mathbf{H}_\mathrm{SI}^H{\mathbf{u}_l}\mathbf{u}_l^H{\mathbf{H}_\mathrm{SI}}} \right\} +\sum\limits_{j \in \mathcal{J}} {{p_j}{{\left| {\mathbf{u}_l^H{\mathbf{f}_j}} \right|}^2}}  + \sigma _\mathrm{BS}^2} }}} .\label{grad_12} 
			\end{align}\hrulefill}
\end{figure*}
Similarly, given the feasible point $\left\{\mathbf{S}_l^c ,\mathbf{W}_k^c,p_j^c \right\}$, the global overestimates of ${\beta _2}\left( {{\mathbf{S}_l},{\mathbf{W}_k},{p_j}} \right)$ and ${\beta _3}\left( {{\mathbf{S}_l},{\mathbf{W}_k}} \right)$ are respectively given by
\begin{align}
	& {\beta _2}\left( {{\mathbf{S}_l},{\mathbf{W}_k},{p_j}} \right) \le {{\hat \beta }_2}\left( {{\mathbf{S}_l},{\mathbf{W}_k},{p_j}\left| {\mathbf{S}_l^c,\mathbf{W}_k^c,p_j^c} \right.} \right) \nonumber\\
	& = {\beta _2}\left( {\mathbf{S}_l^c,\mathbf{W}_k^c,p_j^c} \right) \nonumber\\
	& + \sum\limits_{l \in \mathcal{L}} {\mathrm{Tr}\left\{ {{{\left( {{\nabla _{{\mathbf{S}_l}}}{\beta _2}\left( {\mathbf{S}_l^c,\mathbf{W}_k^c,p_j^c} \right)} \right)}^H}\left( {{\mathbf{S}_l} - \mathbf{S}_l^c} \right)} \right\}} \nonumber\\  
	& + \sum\limits_{k \in \mathcal{K}} {\mathrm{Tr}\left\{ {{{\left( {{\nabla _{{\mathbf{W}_k}}}{\beta _2}\left( {\mathbf{S}_l^c,\mathbf{W}_k^c,p_j^c} \right)} \right)}^H}\left( {{\mathbf{W}_k} - \mathbf{W}_k^c} \right)} \right\}}  \nonumber\\ 
	& + \sum\limits_{j \in \mathcal{J}} {{\nabla _{{p_j}}}{\beta _2}\left( {\mathbf{S}_l^c,\mathbf{W}_k^c,p_j^c} \right)\left( {{p_j} - p_j^c} \right)} ,
\end{align}
and
\begin{align}
	& {\beta _3}\left( {{\mathbf{S}_l},{\mathbf{W}_k}} \right) \le {{\hat \beta }_3}\left( {{\mathbf{S}_l},{\mathbf{W}_k}\left| {\mathbf{S}_l^c,\mathbf{W}_k^c} \right.} \right) \nonumber\\
	& = {\beta _3}\left( {\mathbf{S}_l^c,\mathbf{W}_k^c} \right) \nonumber\\
	& + \sum\limits_{l \in \mathcal{L}} {\mathrm{Tr}\left\{ {{{\left( {{\nabla _{{\mathbf{S}_l}}}{\beta _3}\left( {\mathbf{S}_l^c,\mathbf{W}_k^c} \right)} \right)}^H}\left( {{\mathbf{S}_l} - \mathbf{S}_l^c} \right)} \right\}} \nonumber\\  
	& + \sum\limits_{k \in \mathcal{K}} {\mathrm{Tr}\left\{ {{{\left( {{\nabla _{{\mathbf{W}_k}}}{\beta _3}\left( {\mathbf{S}_l^c,\mathbf{W}_k^c} \right)} \right)}^H}\left( {{\mathbf{W}_k} - \mathbf{W}_k^c} \right)} \right\}},
\end{align}
where gradients ${\nabla _{{\mathbf{S}_l}}}{\beta _2}$, ${\nabla _{{\mathbf{W}_k}}}{\beta _2}$, ${\nabla _{p_j}}{\beta _2}$, ${\nabla _{{\mathbf{S}_l}}}{\beta _3}$, and ${\nabla _{{\mathbf{W}_k}}}{\beta _3}$ are shown in \eqref{grad_23_begin}-\eqref{grad_23_end} at the top of the next page.
\begin{figure*}[!t]
	\textsc{\centering
		\begin{align}
			& {\nabla _{{\mathbf{S}_l}}}{\beta _2} = {\nabla _{{\mathbf{W}_k}}}{\beta _2} = \sum\limits_{j \in \mathcal{J}} {\frac{{{\varpi _{{\mathbf{U}_j}}}}}{{\ln 2}} \cdot \frac{{\sum\limits_{l \in \mathcal{L}} {\mathbf{G}_l^H{\mathbf{b}_j}\mathbf{b}_j^H{\mathbf{G}_l}}  + \mathbf{H}_\mathrm{SI}^H{\mathbf{b}_j}\mathbf{b}_j^H{\mathbf{H}_\mathrm{SI}}}}{{\sum\limits_{l \in \mathcal{L}} {\mathrm{Tr}\left\{ { {\mathbf{R}}\mathbf{G}_l^H{\mathbf{b}_j}\mathbf{b}_j^H{\mathbf{G}_l}} \right\}}  + \mathrm{Tr}\left\{ { {\mathbf{R}}\mathbf{H}_\mathrm{SI}^H{\mathbf{b}_j}\mathbf{b}_j^H{\mathbf{H}_\mathrm{SI}}} \right\} + \sum\limits_{i \in \mathcal{J} \backslash j} {{p_i}{{\left| {\mathbf{b}_j^H{\mathbf{f}_i}} \right|}^2}}  + \sigma _\mathrm{BS}^2}}} .\label{grad_23_begin}\\
			& {\nabla _{{p_j}}}{\beta _2} = \sum\limits_{t \in \mathcal{J}\backslash j} {\frac{{{\varpi _{{\mathrm{U}_t}}}}}{{\ln 2}} \cdot \frac{{{{\left| {\mathbf{b}_t^H{\mathbf{f}_j}} \right|}^2}}}{{\sum\limits_{l \in \mathcal{L}} {\mathrm{Tr}\left\{ { {\mathbf{R}}\mathbf{G}_l^H{\mathbf{b}_t}\mathbf{b}_t^H{\mathbf{G}_l}} \right\}}  + \mathrm{Tr}\left\{ { {\mathbf{R}}\mathbf{H}_\mathrm{SI}^H{\mathbf{b}_t}\mathbf{b}_t^H{\mathbf{H}_\mathrm{SI}}} \right\} + \sum\limits_{i \in \mathcal{J} \backslash t} {{p_i}{{\left| {\mathbf{b}_t^H{\mathbf{f}_i}} \right|}^2}}  + \sigma _\mathrm{BS}^2}}} . \\
			& {\nabla _{{\mathbf{S}_l}}}{\beta _3} = \sum\limits_{k \in \mathcal{K}} {\frac{{{\varpi _{{\mathrm{D}_k}}}}}{{\ln 2}} \cdot \frac{{{\mathbf{h}_k}\mathbf{h}_k^H}}{{\mathrm{Tr}\left\{ {\left( {\sum\limits_{l \in \mathcal{L}} {{\mathbf{S}_l}}  + \sum\limits_{i \in \mathcal{K}\backslash k} {{\mathbf{W}_i}} } \right){\mathbf{h}_k}\mathbf{h}_k^H} \right\} + \sigma _k^2}}} .\\
			& {\nabla _{{\mathbf{W}_k}}}{\beta _3} = \sum\limits_{t \in \mathcal{K}\backslash k} {\frac{{{\varpi _{{\mathrm{D}_t}}}}}{{\ln 2}} \cdot \frac{{{\mathbf{h}_t}\mathbf{h}_t^H}}{{\mathrm{Tr}\left\{ {\left( {\sum\limits_{l \in \mathcal{L}} {{\mathbf{S}_l}}  + \sum\limits_{i \in \mathcal{K}\backslash t} {{\mathbf{W}_i}} } \right){\mathbf{h}_t}\mathbf{h}_t^H} \right\} + \sigma _k^2}}} . \label{grad_23_end}
	\end{align}\hrulefill}
\end{figure*}
Therefore, in the $c$-th iteration, given the feasible point $\left\{\mathbf{S}_l^c ,\mathbf{W}_k^c,p_j^c \right\}$, a lower bound of objective value \eqref{ccv_R} can be determined by 
\begin{align}\label{WSR1}
	\widehat {WSR}\left( {\mathbf{S}_l,\mathbf{W}_k,p_j} \right) & = {\alpha _1} + {\alpha _2} + {\alpha _3} \nonumber\\
	& - {{\hat \beta }_1}\left( {{\mathbf{S}_l},{\mathbf{W}_k},{p_j}\left| {\mathbf{S}_l^c,\mathbf{W}_k^c,p_j^c} \right.} \right) \nonumber\\
	& - {{\hat \beta }_2}\left( {{\mathbf{S}_l},{\mathbf{W}_k},{p_j}\left| {\mathbf{S}_l^c,\mathbf{W}_k^c,p_j^c} \right.} \right) \nonumber\\
	& - {{\hat \beta }_3}\left( {{\mathbf{S}_l},{\mathbf{W}_k}\left| {\mathbf{S}_l^c,\mathbf{W}_k^c} \right.} \right).
\end{align}
Then, problem \eqref{max_sca2} can be reformulated as
\begin{align} \label{max_sca3}
	& \mathop {\mathrm{maximize} }\limits_{\mathbf{S}_l,\mathbf{W}_k,p_j} \quad \widehat {WSR}\left( {\mathbf{S}_l,\mathbf{W}_k,p_j} \right) \\
	&\mathrm{s.t.} \quad \mathrm{C3a},\mathrm{C3b},\mathrm{C3c},\mathrm{C4}. \nonumber
\end{align}
Next, for the rank constraint C3c, we apply semidefinite relaxation (SDR) and remove constraint C3c. The relaxed version of problem \eqref{max_sca3} can now be optimally solved using standard convex solvers such as CVX \cite{add3}. Then, we verify the tightness of SDR in the following proposition.
\begin{theorem}
If $P_\mathrm{D}^\mathrm{max}>0$, the optimal beamforming matrices $\mathbf{W}_k$ satisfying $\mathrm{Rank}\left\{\mathbf{W}_k \right\} \le 1$ can always be obtained.
\end{theorem}
\begin{IEEEproof}
	Please refer to \cite[Appendix A]{rank1}.
\end{IEEEproof}
The SCA algorithm for solving problem \eqref{max_sca1} iteratively is summarized in Algorithm \ref{SCA}.
\begin{algorithm}[!t]
	\caption{SCA algorithm for solving problem \eqref{max_sca1}}
	\label{SCA}
	\renewcommand{\algorithmicrequire}{\textbf{Initialization:}}
	\renewcommand{\algorithmicensure}{\textbf{Output:}}
	\begin{algorithmic}[1]
		\REQUIRE Set initial point $\left\{\mathbf{S}_l^0 ,\mathbf{W}_k^0,p_j^0 \right\}$, iteration index $c = 0$, and error tolerance $0 \le \epsilon \ll 1$.
		\ENSURE ${\mathbf{S}_l}$, ${\mathbf{w}_k}$, and ${p_j}$.
		\STATE Calculate initial value $\widehat {WSR}\left( {\mathbf{S}_l^0,\mathbf{W}_k^0,p_j^0} \right)$;
		\REPEAT
		\STATE Set $c=c+1$;
		\STATE Solve the relaxed version of problem \eqref{max_sca3} for the given feasible point $\left\{\mathbf{S}_l^{c-1} ,\mathbf{W}_k^{c-1},p_j^{c-1} \right\}$ and store the intermediate solution $\left\{\mathbf{S}_l^c ,\mathbf{W}_k^c,p_j^c \right\}=\left\{\mathbf{S}_l ,\mathbf{W}_k,p_j \right\}$;		
		\UNTIL {Increase of objective value \eqref{WSR1} is less than $\epsilon$ or $c \ge C$}
		\STATE Perform eigenvalue decomposition on $\mathbf{W}_k^c$ to obtain $\mathbf{w}_k$;
		\RETURN ${\mathbf{S}_l}={\mathbf{S}_l^c}$, $\mathbf{w}_k$, and $p_j=p_j^c$.
	\end{algorithmic}
\end{algorithm}
\subsubsection{AO Algorithm for Solving Problem \eqref{max2}}
After obtaining the solutions of sub-problems 1 and 2, the proposed AO algorithm for solving problem \eqref{max2} is summarized in Algorithm \ref{AO}. Specifically, we first obtain the optimized $\left\{\mathbf{u}_l,\mathbf{b}_j  \right\}$ by closed-form expressions \eqref{receive_beam1} and \eqref{receive_beam2} (Line 4). Then, we optimize $\left\{\mathbf{S}_l,\mathbf{w}_k,p_j \right\}$ by solving sub-problem 2 based on SCA (Line 5). The AO algorithm iteratively solves the two sub-problems until the increase in objective value \eqref{sum_rate} is less than error tolerance threshold $\tilde \epsilon$ or the maximum number of iterations for AO, $\widetilde C$, is reached.
\begin{algorithm}[!t]
	\caption{AO algorithm for solving problem \eqref{max2}}
	\label{AO}
	\renewcommand{\algorithmicrequire}{\textbf{Initialization:}}
	\renewcommand{\algorithmicensure}{\textbf{Output:}}
	\begin{algorithmic}[1]
		\REQUIRE Set initial ${\left\{ \mathbf{u}_l^0,\mathbf{b}_j^0,\mathbf{S}_l^0,\mathbf{w}_k^0,p_j^0 \right\}}$, iteration index $\tilde c = 0$, and error tolerance $0 \le \tilde \epsilon \ll 1$.
		\ENSURE $\mathbf{u}_l$, $\mathbf{b}_j$, $\mathbf{S}_l$, $\mathbf{w}_k$, and $p_j$.
		\STATE Calculate initial value $WSR\left(\mathbf{u}_l^0,\mathbf{b}_j^0,\mathbf{S}_l^0,\mathbf{w}_k^0,p_j^0 \right) $;
		\REPEAT
		\STATE Set $\tilde c=\tilde c+1$;
		\STATE With given $\mathbf{S}_l^{\tilde c-1}$, $\mathbf{w}_k^{\tilde c-1}$, and $\mathbf{p}_j^{\tilde c-1}$, solve sub-problem 1 by \eqref{receive_beam1} and \eqref{receive_beam2}, and store the intermediate solutions $\mathbf{u}_l^{\tilde c} = \mathbf{u}_l$ and $\mathbf{b}_j^{\tilde c}=\mathbf{b}_j$;
		\STATE With given $\mathbf{u}_l^{\tilde c}$ and $\mathbf{b}_j^{\tilde c}$, solve sub-problem 2 by Algorithm \ref{SCA} and store the intermediate solutions $\mathbf{S}_l^{\tilde c}=\mathbf{S}_l$, $\mathbf{w}_k^{\tilde c}=\mathbf{w}_k$, and $\mathbf{p}_j^{\tilde c}=\mathbf{p}_j$;
		\UNTIL {Increase in objective value \eqref{sum_rate} is less than $\tilde \epsilon$ or $\tilde c \ge \widetilde C$}
		\RETURN $\mathbf{u}_l=\mathbf{u}_l^{\tilde c}$, $\mathbf{b}_j=\mathbf{b}_j^{\tilde c}$, $\mathbf{S}_l=\mathbf{S}_l^{\tilde c}$, $\mathbf{w}_k=\mathbf{w}_k^{\tilde c}$, and $p_j=p_j^{\tilde c}$.
	\end{algorithmic}
\end{algorithm}
\subsection{Outer-Layer of RP Algorithm}
In the outer-layer, we propose an intuitive antenna position optimization algorithm, i.e., RP algorithm, to obtain the optimized positions for multiple transmit and receive MAs. 
\begin{algorithm}[!t]
	\caption{RP algorithm for solving problem \eqref{max1}}
	\label{RP}
	\renewcommand{\algorithmicrequire}{\textbf{Input:}}
	\renewcommand{\algorithmicensure}{\textbf{Output:}}
	\begin{algorithmic}[1]
		\REQUIRE  $M$, $N$, $K$, $J$, $L$, $P^{\max}_\mathrm{D}$, $P^{\max}_\mathrm{U}$, $D$, $\Gamma$, $\mathcal{C}_\mathrm{t}$, $\mathcal{C}_\mathrm{r}$, $\rho_\mathrm{SI}$, $\lambda$, $\left\{ {{\rho_{{\mathrm{D}_k}}}} \right\}$, $\left\{ {{\rho_{{\mathrm{U}_j}}}} \right\}$, $\left\{ {{\rho_{{\mathrm{S}_l}}}} \right\}$, $\left\{ {{\mathbf{q}_{{\mathrm{U}_j}}}} \right\}$, $\left\{ {{\mathbf{q}_{{\mathrm{D}_k}}}} \right\}$, and $\left\{ {{\mathbf{q}_{{\mathrm{S}_l}}}} \right\}$.
		\ENSURE $\mathbf{u}_l$, $\mathbf{b}_j$, $\mathbf{S}_l$, $\mathbf{w}_k$, $p_j$, $\mathbf{t}$, and $\mathbf{r}$.
		\STATE Randomly generate $\left\{ {{{\left\{ {\mathbf{t},\mathbf{r}} \right\}}_i}} \right\}_{i = 1}^\Gamma $ that satisfy constraints C5-C7;
		\FOR{$i=1:1:\Gamma$}
		\STATE With given ${{{\left\{ {\mathbf{t},\mathbf{r}} \right\}}_i}}$, solve problem \eqref{max2} by Algorithm \ref{AO} to obtain the optimized $\left\{\mathbf{u}_l, \mathbf{b}_j, \mathbf{S}_l, \mathbf{w}_k, p_j\right\}$;
		\STATE Calculate the corresponding $WSR\left( {{{\left\{ {\mathbf{t},\mathbf{r}} \right\}}_i}} \right)$ by \eqref{sum_rate};
		\ENDFOR
		\STATE Select the optimized MA position $\left\{ {\mathbf{t},\mathbf{r}} \right\} = \mathop {\arg \max }\limits_{{{\left\{ {\mathbf{t},\mathbf{r}} \right\}}_i}} \left\{ {WSR\left( {{{\left\{ {\mathbf{t},\mathbf{r}} \right\}}_1}} \right), \ldots ,WSR\left( {{{\left\{ {\mathbf{t},\mathbf{r}} \right\}}_\Gamma }} \right)} \right\}$
		\RETURN $\mathbf{u}_l$, $\mathbf{b}_j$, $\mathbf{S}_l$, $\mathbf{w}_k$, $p_j$, $\mathbf{t}$, and $\mathbf{r}$.
	\end{algorithmic}
\end{algorithm}
It is well known that the adjustment of MA positions effectively reconfigures the channel \cite{MA_zhu2}. In addition, the optimization of beamformers, sensing signal covariance matrices, and UL power allocation relies directly on the channel response. Therefore, in the antenna position optimization process, the beamformers, sensing signal covariance matrices, and UL power allocation need to be optimized for each candidate MA position to fully exploit the performance potential of the current MA position. To reduce the computational complexity of the antenna position optimization algorithm, we provide an intuitive and feasible RP algorithm, which is summarized in Algorithm \ref{RP}. First, the RP algorithm randomly generate $\Gamma$ pairs of $\mathbf{t}$ and $\mathbf{r}$, i.e., $\left\{\mathbf{t},\mathbf{r}\right\}_i$ ($1 \le i \le \Gamma$), that satisfy constraints C5-C7 (Line 1). For each pair, problem \eqref{max2} is solved by Algorithm \ref{AO} to obtain the optimized $\left\{\mathbf{u}_l, \mathbf{b}_j, \mathbf{S}_l, \mathbf{w}_k, p_j\right\}$ (Line 3). The corresponding WSR is calculated based on \eqref{sum_rate} (Line 4). Then, the $\left\{\mathbf{t},\mathbf{r}\right\}$ with the largest WSR is selected as the optimized MA position (Line 6). Finally, the optimized MA position, along with the corresponding beamformers, sensing signal covariance matrices, and UL power allocation, are obtained.
\subsection{Antenna Position Matching (APM) Algorithm}
To reduce the additional overhead caused by antenna movement over large-size regions, we develop the APM algorithm to minimize the total MA movement distance in this sub-section.
\begin{figure}[!t]
	\centering
	\includegraphics[width=1\linewidth]{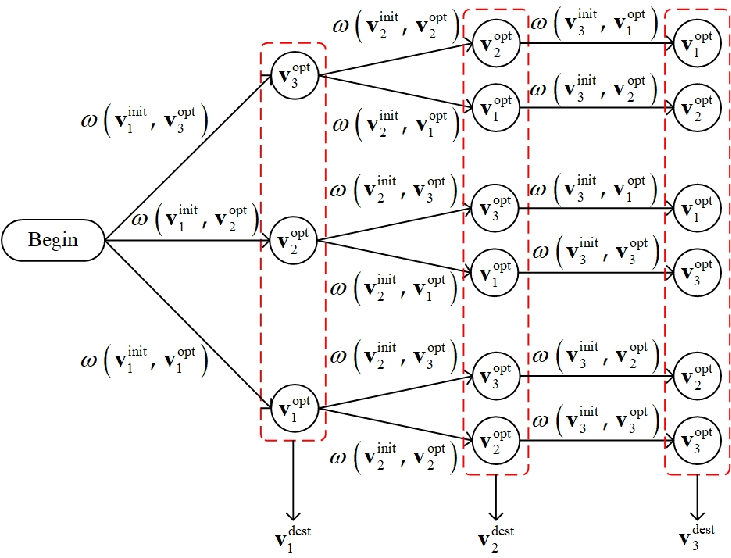}
	\caption{An example of APM with $\widetilde{N}=3$.}
	\label{MA_match}
\end{figure}

Define the initial and optimized antenna positions of $\widetilde{N}$ MAs before and after antenna position optimization as $\mathbf{v}_{\tilde{n}}^{\mathrm{init}} \in \mathbb{R}^{3 \times 1}$ and $\mathbf{v}_{\tilde{n}}^{\mathrm{opt}} \in \mathbb{R}^{3 \times 1}$ ($1 \le \tilde{n} \le \widetilde{N}$), respectively. Prior works on MA-aided systems have yet incorporated the MA movement distance into the objective function or constraints of the optimization problem, apart from an initial investigation given in \cite{MA_delay}. In other words, when moving MA $\tilde{n}$ from $\mathbf{v}_{\tilde{n}}^\mathrm{init}$ to $\mathbf{v}_{\tilde{n}}^\mathrm{opt}$, they do not consider the additional costs incurred by antenna movement in practical applications. The optimized position $\mathbf{v}_{\tilde{n}}^\mathrm{opt}$ may be far from the corresponding initial position $\mathbf{v}_{\tilde{n}}^\mathrm{init}$ but closer to another initial MA position. Furthermore, this oversight is especially significant in near-field ISAC scenarios because the large-size moving regions are deployed. As such, we propose an APM algorithm based on the greedy strategy to tackle this issue. 

The total MA movement distance minimization problem is a shortest path search problem in tree graph. Fig. \ref{MA_match} shows an example for the case when $\widetilde{N}=3$, where 
\begin{equation} \label{path_weight}
	\omega \left( {\mathbf{v}_{\tilde{n}}^\mathrm{init},\mathbf{v}_{\hat n}^\mathrm{opt}} \right) = {\left\| {\mathbf{v}_{\tilde{n}}^\mathrm{init} - \mathbf{v}_{\hat n}^\mathrm{opt}} \right\|_2}, \quad 1\le {\hat n} \le \widetilde{N},
\end{equation}
is the path weight determined by the movement distance from $\mathbf{v}_{\tilde{n}}^\mathrm{init}$ to $\mathbf{v}_{\hat n}^\mathrm{opt}$. For each MA, the destination position, $\mathbf{v}_{\tilde{n}}^\mathrm{dest} \in \mathbb{R}^{3 \times 1}$, can be selected according to path weights. Once a specific optimized position is selected by one MA, it becomes unavailable for the remaining MAs. Consequently, there are $\widetilde{N}!$ APM solutions.

To find the optimal APM solution that minimizes the total MA movement distance, a straightforward approach is to perform an exhaustive search over the $\widetilde{N}!$ solutions. However, this clearly results in high computational complexity, especially when $\widetilde{N}$ is large. As a result, we adopt the greedy strategy to identify a suboptimal solution. First, we initialize the antenna index set as $\widetilde {\mathcal{N}}^0 = \left\{ {1, \ldots ,\widetilde {N}} \right\}$. Next, we sequentially select destination position $\mathbf{v}_{\tilde{n}}^\mathrm{dest}$ for each MA. Specifically, for MA $\tilde{n}$, we select the optimized position in updated antenna index set $\widetilde {\mathcal{N}}^{\tilde{n}-1}$ with the smallest path weight as its destination position, i.e., 
\begin{equation} \label{select_opt}
	\mathbf{v}_{\tilde{n}}^\mathrm{dest} = \mathop {\arg \min }\limits_{\mathbf{v}_{\hat n}^\mathrm{opt}} \left\{ {\omega \left( {\mathbf{v}_{\tilde{n}}^\mathrm{init},\mathbf{v}_{\hat n}^\mathrm{opt}} \right)\left| {{\hat n} \in {{\widetilde {\mathcal{N}}}^{\tilde{n}-1}}} \right.} \right\}.
\end{equation}
Then, we update the antenna index set by removing the index of the selected antenna position, i.e.,
\begin{equation} \label{update_set}
	{\widetilde {\mathcal{N}}}^{\tilde{n}} = {\widetilde {\mathcal{N}}}^{\tilde{n}-1}\backslash {\hat n}.
\end{equation}
After the destination positions for all $\widetilde N$ MAs have been selected, the APM solution $\mathbf{v}_{\tilde{n}}^\mathrm{dest}$ for $1 \le \tilde{n} \le \widetilde{N}$ is obtained.

\begin{algorithm}[!t]
	\caption{APM algorithm for minimizing the total MA movement distance}
	\label{MA_match_alg}
	\renewcommand{\algorithmicrequire}{\textbf{Input:}}
	\renewcommand{\algorithmicensure}{\textbf{Output:}}
	\begin{algorithmic}[1]
		\REQUIRE  $M$, $N$, $\mathbf{t}^{\mathrm{init}}$, $\mathbf{r}^{\mathrm{init}}$, $\mathbf{t}^{\mathrm{opt}}$, and $\mathbf{r}^{\mathrm{opt}}$.
		\ENSURE $\mathbf{t}^{\mathrm{dest}}$ and $\mathbf{r}^{\mathrm{dest}}$.
		\STATE Initialize the index sets of the transmit and receive MAs as $\mathcal{N}^0 = \left\{ {1, \ldots ,N} \right\}$ and $\mathcal{M}^0 = \left\{ {1, \ldots ,M} \right\}$, respectively;
		\FOR{$n=1:1:N$}
		\STATE Calculate path weight $\omega \left( {\mathbf{t}_n^\mathrm{init},\mathbf{t}_{\hat n}^\mathrm{opt}} \right)$ according to \eqref{path_weight};
		\STATE Select optimized position $\mathbf{t}_{\hat n}^\mathrm{opt}$ as transmit MA $n$'s destination position $\mathbf{t}_n^\mathrm{dest}$ according to \eqref{select_opt};
		\STATE Update antenna index set $\mathcal{N}^n$ according to \eqref{update_set};
		\ENDFOR
		\FOR{$m=1:1:M$}
		\STATE Calculate path weight $\omega \left( {\mathbf{r}_m^\mathrm{init},\mathbf{r}_{\hat m}^\mathrm{opt}} \right)$ according to \eqref{path_weight};
		\STATE Select the optimized position $\mathbf{r}_{\hat m}^\mathrm{opt}$ as receive MA $m$'s destination position $\mathbf{r}_m^\mathrm{dest}$ according to \eqref{select_opt};
		\STATE Update antenna index set $\mathcal{M}^m$ according to \eqref{update_set};
		\ENDFOR
		\RETURN $\mathbf{t}^{\mathrm{dest}}$ and $\mathbf{r}^{\mathrm{dest}}$.
	\end{algorithmic}
\end{algorithm}
Note that the aforementioned APM algorithm can be widely applied in MA-aided communication systems. For the proposed system, we need to perform APM separately for the transmit and receive MAs. In other words, we have $\widetilde{N}=N$ or $M$, $\mathbf{v}_{\tilde{n}}^{\mathrm{init}} =\mathbf{t}_n^{\mathrm{init}}$ or $\mathbf{r}_m^{\mathrm{init}}$, and $\mathbf{v}_{\tilde{n}}^{\mathrm{opt}}=\mathbf{t}_n^{\mathrm{opt}}$ or $\mathbf{r}_m^{\mathrm{opt}}$, where $\mathbf{t}_n^{\mathrm{init}}/\mathbf{r}_m^{\mathrm{init}}$ is the initial transmit/receive MA position and $\mathbf{t}_n^{\mathrm{opt}}/\mathbf{r}_m^{\mathrm{opt}}$ is the optimized transmit/receive MA position output by Algorithm \ref{RP}. The corresponding processing steps are summarized in Algorithm \ref{MA_match_alg}. After that, the destination positions of transmit and receive MAs, $\mathbf{t}_n^{\mathrm{dest}}$ and $\mathbf{r}_m^{\mathrm{dest}}$, can be obtained.

In practical applications, the movement of multiple MAs within a given region can be realized through two typical methods \cite{MA_app}. The first method divides the moving region into several non-overlapping sub-regions, where each MA is restricted to move only within its designated sub-region. 
The second method is based on dense array antennas. Specifically, a large number of antennas are compactly arranged within the transmit/receive region, and each is integrated with a reconfigurable device (e.g., a pixel antenna \cite{pixel}) to enable dynamic configuration. By adjusting the states of these reconfigurable devices, different sets of antennas can be selectively activated. This method serves as an efficient alternative to physical MA movement and is particularly well-suited for scenarios with limited movement space.
\subsection{Overall Algorithm}
\begin{algorithm}[!t]
	\caption{Overall algorithm for solving problem \eqref{max1} and minimizing total MA movement distance}
	\label{overall}
	\renewcommand{\algorithmicrequire}{\textbf{Input:}}
	\renewcommand{\algorithmicensure}{\textbf{Output:}}
	\begin{algorithmic}[1]
		\REQUIRE  $M$, $N$, $K$, $J$, $L$, $P^{\max}_\mathrm{D}$, $P^{\max}_\mathrm{U}$, $D$, $\Gamma$, $\mathcal{C}_\mathrm{t}$, $\mathcal{C}_\mathrm{r}$, $\rho_\mathrm{SI}$, $\lambda$, $\left\{ {{\rho_{{\mathrm{D}_k}}}} \right\}$, $\left\{ {{\rho_{{\mathrm{U}_j}}}} \right\}$, $\left\{ {{\rho_{{\mathrm{S}_l}}}} \right\}$, $\left\{ {{\mathbf{q}_{{\mathrm{U}_j}}}} \right\}$, $\left\{ {{\mathbf{q}_{{\mathrm{D}_k}}}} \right\}$, $\left\{ {{\mathbf{q}_{{\mathrm{S}_l}}}} \right\}$, $\mathbf{t}^{\mathrm{init}}$, and $\mathbf{r}^{\mathrm{init}}$.
		\ENSURE $\mathbf{u}_l$, $\mathbf{b}_j$, $\mathbf{S}_l$, $\mathbf{w}_k$, $p_j$, $\mathbf{t}^{\mathrm{dest}}$, and $\mathbf{r}^{\mathrm{dest}}$.
		\STATE Solving problem \eqref{max1} by Algorithm \ref{RP} to obtain optimized $\mathbf{u}_l$, $\mathbf{b}_j$, $\mathbf{S}_l$, $\mathbf{w}_k$, $p_j$, $\mathbf{t}^{\mathrm{opt}}$, and $\mathbf{r}^{\mathrm{opt}}$;
		\STATE Search destination positions for transmit and receive MAs, $\mathbf{t}^{\mathrm{dest}}$ and $\mathbf{r}^{\mathrm{dest}}$, by Algorithm \ref{MA_match_alg};
		\RETURN $\mathbf{u}_l$, $\mathbf{b}_j$, $\mathbf{S}_l$, $\mathbf{w}_k$, $p_j$, $\mathbf{t}^{\mathrm{dest}}$, and $\mathbf{r}^{\mathrm{dest}}$.
	\end{algorithmic}
\end{algorithm}
The detailed overall algorithm for solving problem \eqref{max1} and minimizing the total MA movement distance is summarized in Algorithm \ref{overall}. Specifically, the optimized transmit beamformers, $\mathbf{w}_k$, sensing signal covariance matrices, $\mathbf{S}_l$, receive beamformers, $\mathbf{u}_l$ and $\mathbf{b}_j$, UL power allocation, $p_j$, and MA positions, $\mathbf{t}^{\mathrm{opt}}$ and $\mathbf{r}^{\mathrm{opt}}$, are obtained by Algorithm \ref{RP} (Line 1). Subsequently, the destination positions of transmit and receive MAs, $\mathbf{t}^{\mathrm{dest}}$ and $\mathbf{r}^{\mathrm{dest}}$, are searched by Algorithm \ref{MA_match_alg} to minimize the total MA movement distance (Line 2). 
\subsection{Convergence and Complexity Analysis}\label{conv_com}
As the RP algorithm only executes the selection of candidate MA positions, its convergence depends on Algorithm \ref{AO}. The convergence of Algorithm \ref{AO} is guaranteed by the following inequality:
\begin{align}
	& WSR\left(\mathbf{u}_l^{\tilde c},\mathbf{b}_j^{\tilde c},\mathbf{S}_l^{\tilde c},\mathbf{w}_k^{\tilde c},p_j^{\tilde c} \right) \nonumber\\
	& \mathop  \ge \limits^{\left( {{\alpha _1}} \right)}   WSR\left(\mathbf{u}_l^{\tilde c},\mathbf{b}_j^{\tilde c},\mathbf{S}_l^{\tilde c-1},\mathbf{w}_k^{\tilde c-1},p_j^{\tilde c-1} \right)	\nonumber\\
	& \mathop  \ge \limits^{\left( {{\alpha _2}} \right)}  WSR\left(\mathbf{u}_l^{\tilde c-1},\mathbf{b}_j^{\tilde c-1},\mathbf{S}_l^{\tilde c-1},\mathbf{w}_k^{\tilde c-1},p_j^{\tilde c-1} \right),
\end{align}
where inequality ${\left( {{\alpha _1}} \right)}$ holds because $\left\{\mathbf{S}_l^{\tilde c},\mathbf{w}_k^{\tilde c},p_j^{\tilde c}\right\}$ are the optimized sensing signal covariance matrices, transmit beamformers, and UL power allocation via Algorithm \ref{SCA} under the current $\left\{\mathbf{u}_l^{\tilde c},\mathbf{b}_j^{\tilde c}\right\}$, and inequality ${\left( {{\alpha _2}} \right)}$ holds because $\left\{\mathbf{u}_l^{\tilde c},\mathbf{b}_j^{\tilde c}\right\}$ are the optimal receive beamformers for maximizing SINRs \eqref{SINR_Sl} and \eqref{SINR_Uj} under the current $\left\{\mathbf{S}_l^{\tilde c-1},\mathbf{w}_k^{\tilde c-1},p_j^{\tilde c-1}\right\}$. Thus, the objective value is non-decreasing during the iterations in Algorithm \ref{AO}. Meanwhile, the objective value is upper-bounded due to finite communication resources. As such, the convergence of Algorithm \ref{AO} is guaranteed. Moreover, the convergence is verified by the simulations in Section \ref{con}.

The main computational complexity of the overall algorithm, i.e., Algorithm \ref{overall}, is due to the iteration in Algorithm \ref{AO}, the selection in Algorithm \ref{RP}, and the search in Algorithm \ref{MA_match_alg}. In Algorithm \ref{AO}, the computational complexity for calculating receive beamformers is $\mathcal{O}\left(\left( L+J\right) M^3 \right) $ due to the matrix inversion in \eqref{receive_beam1} and \eqref{receive_beam2}. The computational complexity of Algorithm \ref{SCA} for optimizing transmit beamformers, sensing signal covariance matrices, and UL power allocation is $\mathcal{O}\left(C_{\mathrm{SCA}}\left( \left( L+K\right)N^{3.5}+J^{3.5} \right) \right) $ due to solving the SDR problem iteratively, where $C_{\mathrm{SCA}}$ is the number of iterations for SCA. Therefore, the computational complexity of Algorithm \ref{AO} is $o_1=\mathcal{O}\left( \widetilde{C}_{\mathrm{AO}}\left(\left( L+J\right) M^3+C_{\mathrm{SCA}}\left( \left( L+K\right)N^{3.5}+J^{3.5} \right) \right) \right)$, where $\widetilde{C}_{\mathrm{AO}}$ is the number of iterations for AO. In addition, the computational complexity of Algorithm \ref{MA_match_alg} for matching antenna positions is $o_2=\mathcal{O}\left( {\frac{1}{2}\left( {N\left( {N + 1} \right) + M\left( {M + 1} \right)} \right)} \right)$. As a result, with the number of candidate MA position pairs $\Gamma$, the computational complexity of the overall algorithm is $\mathcal{O}\left(\Gamma o_1+o_2 \right) $.

\section{Performance Evaluation}\label{4}
In this section, we provide the simulation results to evaluate the performance of the proposed MA-aided near-field ISAC system. First, the simulation setup is introduced, and then the numerical results are presented.
\subsection{Simulation Setup}
\begin{table}[!t]
	\renewcommand{\arraystretch}{1.5}
	\caption{Simulation Parameters}
	\label{tab1}
	\centering
	\begin{tabular}{|l|l|l|}
		\hline
		\multicolumn{1}{|c|}{\textbf{Parameter}} & \multicolumn{1}{c|}{\textbf{Description}} & \multicolumn{1}{c|}{\textbf{Value}} \\ \hline
		$A$ & Side length of moving region & $100\lambda$ \\ \hline
		$N$, $M$ & Number of antennas & 8 \\ \hline
		$L$, $J$, $K$ & Number of targets/users & 2 \\ \hline
		$C$, $\widetilde{C}$ & Maximum number of iterations & 100   \\ \hline
		$\Gamma$ & Number of MA position pairs & 100   \\ \hline
		$\epsilon$, $\tilde \epsilon$ & Error tolerance  & $10^{-3}$   \\ \hline
		$\rho_\mathrm{SI}$& SI loss coefficient  & $-100$ dB   \\ \hline
		$\rho_{\mathrm{S}_l}$ & Round-trip channel coefficient & $-50$ dB   \\ \hline
		$P_{\mathrm{U}}^{\mathrm{max}}$ & Maximum UL transmit power & 10 dBm   \\ \hline
		$P_{\mathrm{D}}^{\mathrm{max}}$ & Maximum DL transmit power & 40 dBm   \\ \hline
		$D$ & Minimum inter-MA distance & $\lambda/2$   \\ \hline
		$\sigma^2_{\mathrm{BS}}$, $\sigma^2_k$ & Average noise power & $-70$ dBm   \\ \hline
	\end{tabular}
\end{table}
\begin{figure}[!t]
	\centering
	\includegraphics[width=1\linewidth]{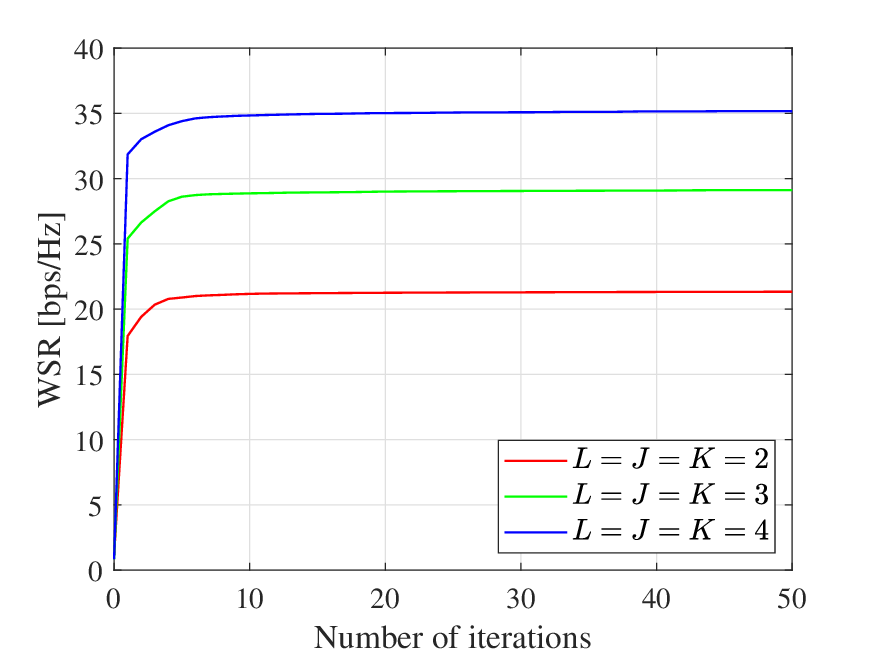}
	\caption{Convergence evaluation of Algorithm \ref{AO}.}
	\label{conv}
\end{figure}
In the simulation, the sensing targets and users are randomly distributed within a semicircular region on the ground centered at the BS, with horizontal distances from the BS to the targets/users ranging from 25 to 30 meters (m). The transmit and receive MA arrays are horizontally mounted on a full-duplex BS at a height of 15 meters. The carrier frequency is set to 30 GHz ($\lambda=0.01$ m). The pass loss coefficients $\rho_{{\mathrm{D}_k}}$ and $\rho_{{\mathrm{U}_j}}$ are determined by the USW model \cite{near3}. Without loss of generality, we assume that the rate weights are equal, i.e., ${\varpi _{{\mathrm{S}_l}}}={\varpi _{{\mathrm{U}_j}}}={\varpi _{{\mathrm{D}_k}}}=\frac{1}{L+J+K}$. Unless otherwise specified, the default simulation parameters are listed in Table\;\ref{tab1}. 
\subsection{Convergence Evaluation of Algorithm \ref{AO}} \label{con}
We first evaluate in Fig. \ref{conv} the convergence performance of the proposed Algorithm \ref{AO}, under different numbers of targets/users. As can be observed, the algorithm demonstrates rapid convergence in all scenarios, with the objective value stabilizing within 10 iterations. This confirms the previous discussion regarding the convergence of Algorithm \ref{AO} in Section \ref{conv_com}. Moreover, the WSR improves as the number of targets/users increases, since the additional targets/users can leverage the redundant spatial DoFs to further enhance the overall system performance.
\subsection{Beamfocusing in MA-Aided Near-Field ISAC}
\begin{figure*}
	\centering
	\subfloat[$A=100\lambda$]{\label{bp_lam1}\includegraphics[width=0.66\columnwidth]{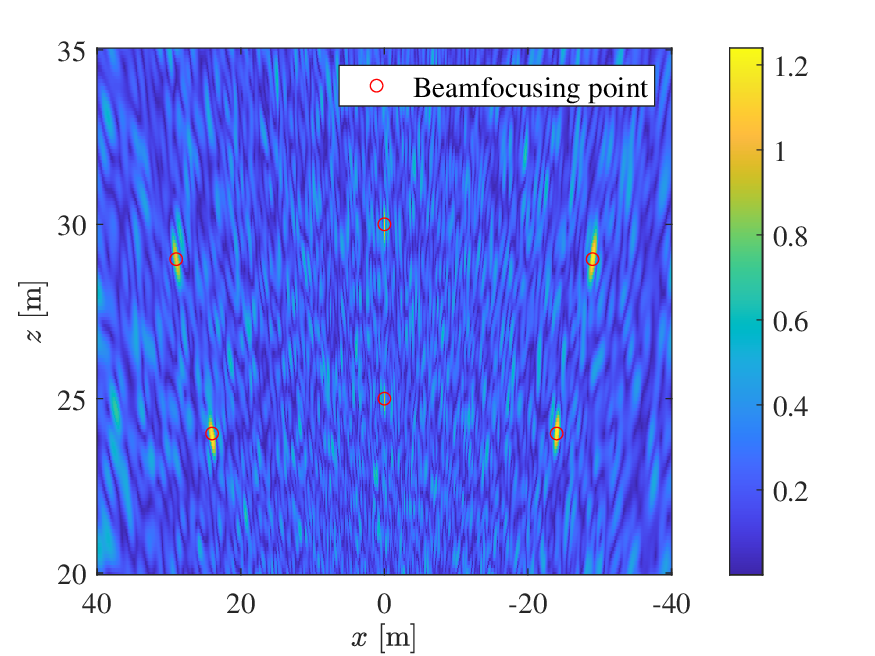}}
	\subfloat[$A=30\lambda$]{\label{bp_lam2}\includegraphics[width=0.66\columnwidth]{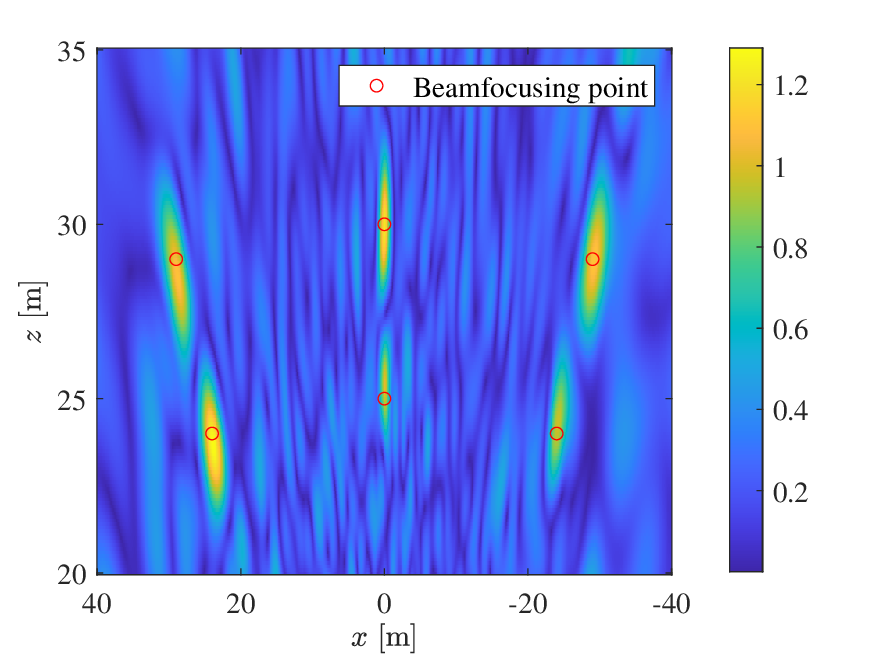}}
	\subfloat[$A=5\lambda$]{\label{bp_lam3}\includegraphics[width=0.66\columnwidth]{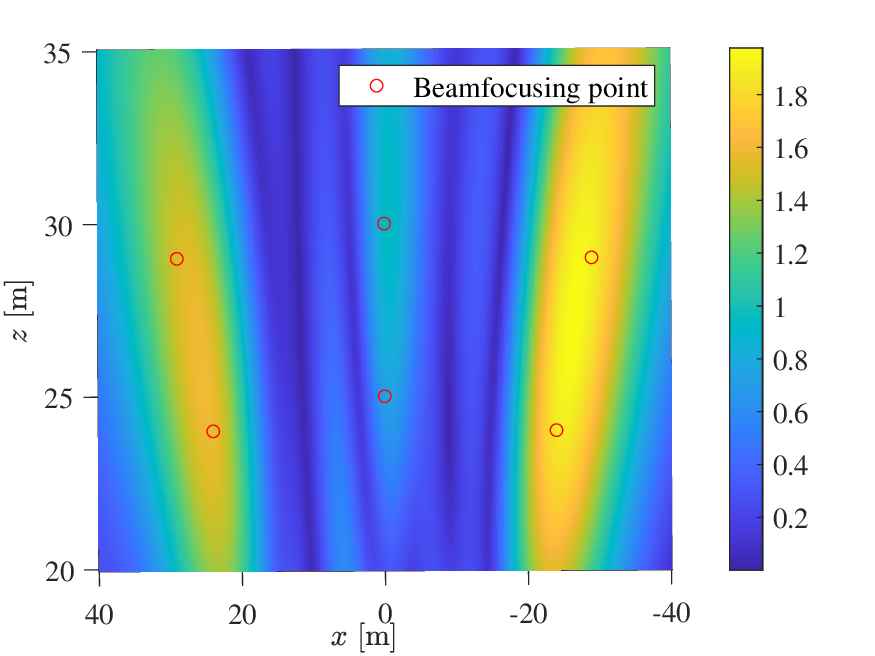}}\\
	\caption{Beampattern under different sizes of the moving region.}
	\label{bp_lam}
\end{figure*}
\begin{figure*}
	\centering
	\subfloat[{[25, 30] m}]{\label{bp_dis1}\includegraphics[width=0.66\columnwidth]{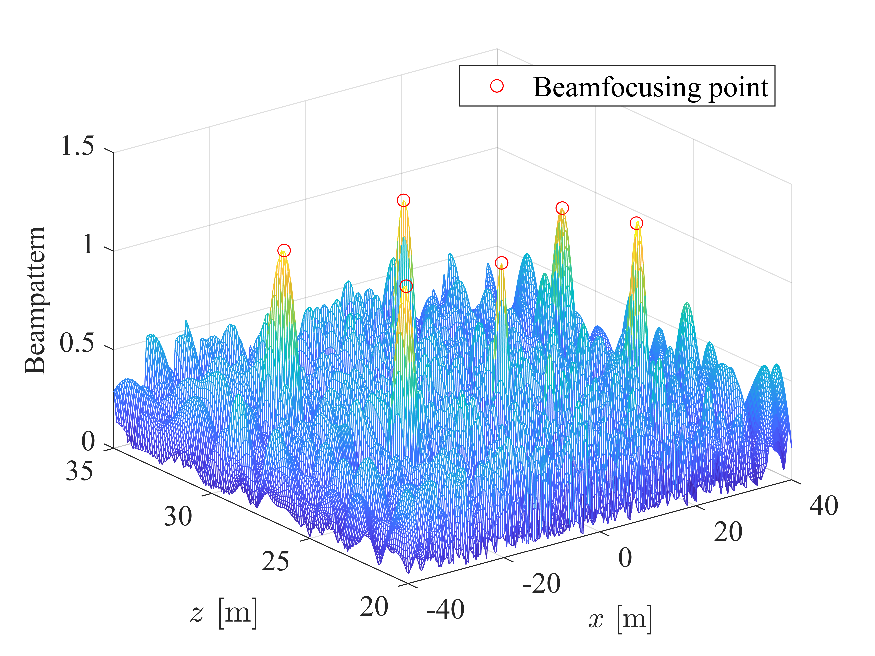}}
	\subfloat[{[95, 100] m}]{\label{bp_dis2}\includegraphics[width=0.66\columnwidth]{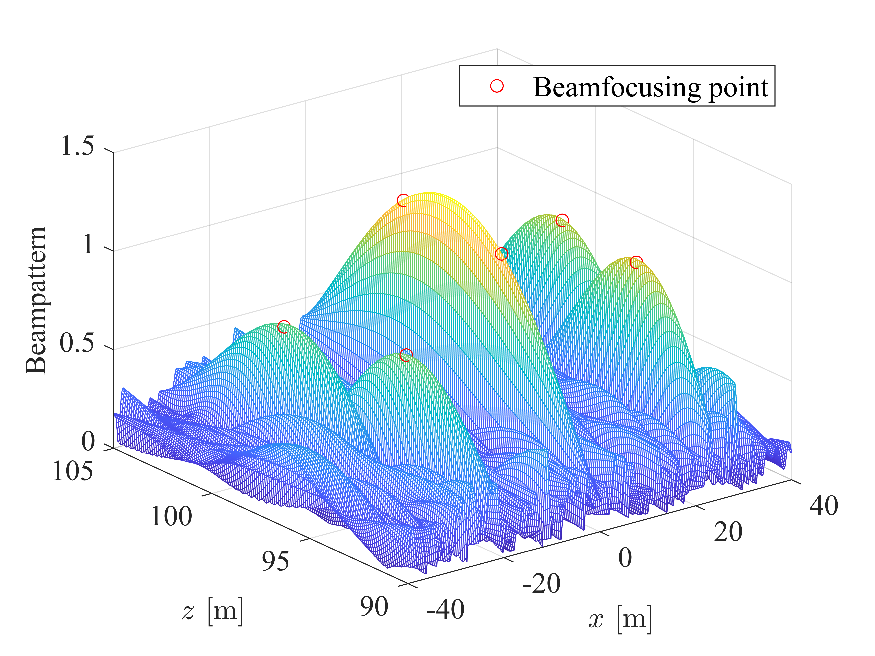}}
	\subfloat[{[495, 500] m}]{\label{bp_dis3}\includegraphics[width=0.66\columnwidth]{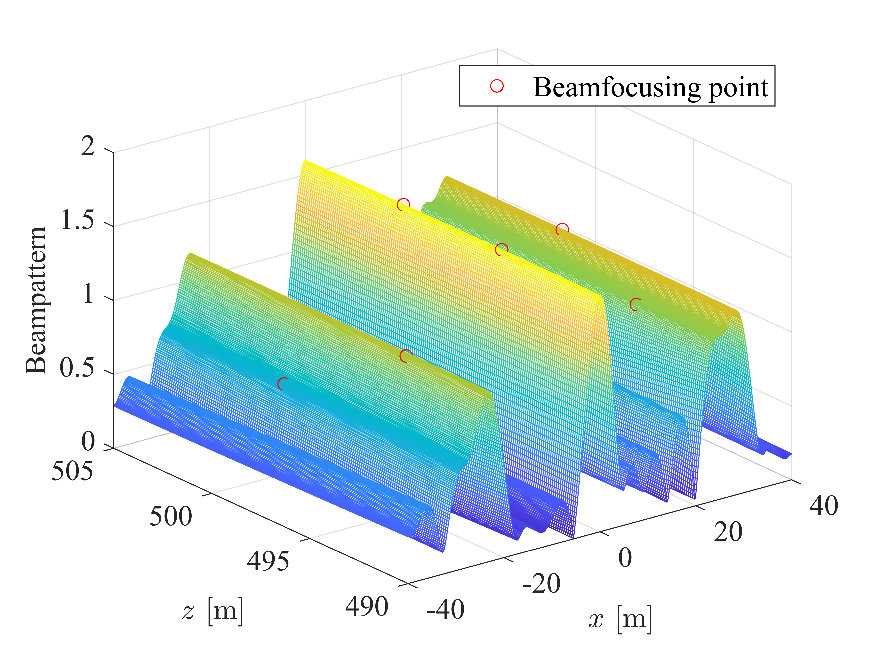}}\\
	\caption{Beampattern under different distances between the BS and beamfocusing points.}
	\label{bp_dis}
\end{figure*}
To intuitively demonstrate the advantages of large-size moving regions in near-field ISAC systems, we present the beampattern corresponding to the optimized geometry of the transmit MA array with $N=128$ in Figs. \ref{bp_lam} and \ref{bp_dis}. The beampattern for the receive MA array can be similarly obtained and is thus omitted here for brevity. Specifically, we calculate the array response vector
\begin{equation}
	\mathbf{a}\left( \mathbf{q} \right) = {\left[ {{e^{\mathrm{j}\frac{{2\pi}}{\lambda }{{\left\| {{\mathbf{t}_1} - \mathbf{q}} \right\|}_2}}}, \ldots ,{e^{\mathrm{j}\frac{{2\pi}}{\lambda }{{\left\| {{\mathbf{t}_N} - \mathbf{q}} \right\|}_2}}}} \right]^T} \in {\mathbb{C}^{N \times 1}},
\end{equation}
for all positions $\mathbf{q}$ within a defined rectangular region on the ground. Then, the beamfocusing at $B$ locations, i.e., ${\tilde {\mathbf{q}}}_b \in {\mathbb{R}^{3 \times 1}}$, $b \in \mathcal{B}=\left\{1,\ldots,B \right\}$, is achieved by setting the beamforming vector as the sum of the array response vectors at these beamfocusing points, i.e.,
\begin{align}
	{\mathbf{w}_\mathrm{bf}} & = \sum\limits_{b \in \mathcal{B}} {\mathbf{a}\left( {{{\tilde {\mathbf{q}}}_b}} \right)}  \nonumber\\
	& = \sum\limits_{b \in \mathcal{B}} {{{\left[ {{e^{\mathrm{j}\frac{{2\pi}}{\lambda }{{\left\| {{\mathbf{t}_1} - {{\tilde {\mathbf{q}}}_b}} \right\|}_2}}}, \ldots ,{e^{\mathrm{j}\frac{{2\pi}}{\lambda }{{\left\| {{\mathbf{t}_N} - {{\tilde {\mathbf{q}}}_b}} \right\|}_2}}}} \right]}^T}}  \in {\mathbb{C}^{N \times 1}},
\end{align}
Therefore, the beamforming gain at position $\mathbf{q}$ can be calculated as
\begin{equation}
	G\left( \mathbf{q} \right) = \frac{1}{N}\left| {\mathbf{w}_\mathrm{bf}^H\mathbf{a}\left( \mathbf{q} \right)} \right|.
\end{equation}

As shown in Fig. \ref{bp_lam}, when $A=100\lambda$ (see Fig. \ref{bp_lam}\subref{bp_dis1}), the beam is precisely focused at the multiple desired locations, and the main lobes of the beam are extremely narrow. This indicates that the beam focused on the desired locations causes minimal interference leakage to other locations, allowing the near-field ISAC system to achieve excellent performance in multi-target sensing and multi-user communication through joint resolutions in both the distance and angle domains. As the size of the moving region decreases ((see Figs. \ref{bp_lam}\subref{bp_dis2} and \ref{bp_lam}\subref{bp_dis3}), the beam's main lobes become wider due to the reduced maximum aperture achievable by the MA array. This results in significant interference to the targets/users at undesired locations, thereby degrading the overall system performance. Notably, when $A=5\lambda$, the beamfocusing points are already in the far-field region, resulting in a loss of resolution in the distance domain and retaining only angle discrimination. In other words, when multiple targets and users are located in the same direction relative to the BS, the far-field ISAC system cannot provide effective sensing and communication services for them. 

In addition, Fig. \ref{bp_dis} provides a 3D perspective on the variations in beampatterns across different distance intervals between the BS and beamfocusing points. The beamforming gains in Fig. \ref{bp_dis}\subref{bp_dis1} at the beamfocusing points are significantly higher compared to the other locations. As the distance between the BS and beamfocusing points increases, the interference leakage from the beam's main lobes to locations that are not beamfocusing points becomes more pronounced because the spherical wave approximates more closely to the plane wave, leading to a gradual disappearance of near-field beamfocusing characteristics. The beamforming gains in Fig. \ref{bp_dis}\subref{bp_dis3} are nearly uniform in the same direction, which makes it impossible to distinguish between different beamfocusing points along that direction. 

Overall, the additional distance dimension of near-field ISAC, compared to far-field ISAC, allows for beamfocusing that enhances the performance of multi-target sensing and multi-user communication. The MA system, which has a larger aperture size of the antenna array compared to conventional FPA systems, naturally expands the near-field region. As a result, the MA-aided near-field ISAC system can benefit from the enlarged near-field region achieved by the large moving region.
\subsection{Performance Comparison with Benchmark Schemes}
To gain more insight, we compare the performance of the proposed scheme (labeled as MA) with benchmark schemes. The considered benchmark schemes for setting antennas’ positions are listed as follows: 1) FPA with full aperture (FPAF): The antenna's positions, $\mathbf{t}$ and $\mathbf{r}$, are set according to the transmit and receive uniform planar arrays (UPAs) with the largest achievable apertures $A \times A$; and 2) FPA with half-wavelength antenna spacing (FPAH): The antenna's positions, $\mathbf{t}$ and $\mathbf{r}$, are set according to the transmit and receive UPAs with half-wavelength inter-antenna spacing both horizontally and vertically. To ensure a fair comparison, the FPAF, FPAH, and MA schemes use the same number of transmit and receive antennas. For the FPAH scheme, the targets and users are in the far-field region of the BS, which facilitates comparing the performance differences between far-field and near-field ISAC systems.

\begin{figure}[!t]
	\centering
	\includegraphics[width=1\linewidth]{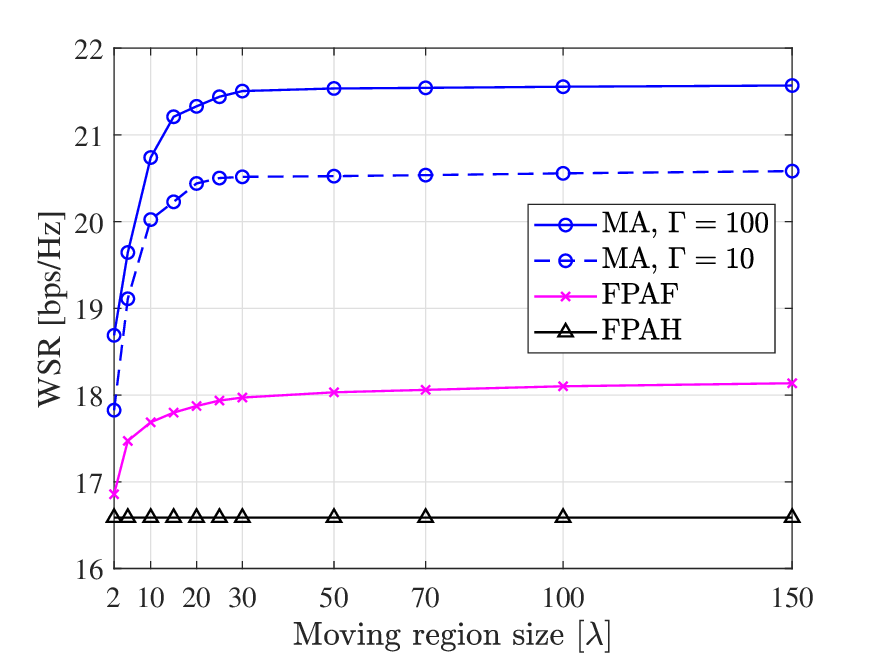}
	\caption{Comparison of WSR between the proposed and benchmark schemes w.r.t. the moving region size.}
	\label{move_size}
\end{figure}
Fig. \ref{move_size} compares the WSRs of different schemes versus the moving region size. As can be seen, the WSRs of the MA and FPAF schemes continuously increase when the moving region size is less than $30\lambda$, and stabilize when the size exceeds $30\lambda$. This is because the increase in the moving region size provides two advantages: 1) expand the optimization space for the antenna position optimization; and 2) enlarge the equivalent array aperture, thereby extending the near-field region, within which the resolution for multiple locations can be achieved. The MA scheme fully leverages both advantages to improve system performance. The FPAF scheme benefits only from advantage 2). The FPAH scheme fails to capitalize on either of these two advantages due to two main limitations. First, the static antenna placement prevents the ISAC system from flexibly optimizing the antenna positions based on varying channel conditions. Second, the fixed array aperture confines the ISAC system to far-field propagation conditions, where it offers resolution only in the angular domain rather than in the distance domain. As a result, the MA scheme achieves a 13.57\% or 19.07\% WSR gain over the FPAF scheme when $\Gamma=10$ or $\Gamma=100$, respectively. The WSR of the FPAH scheme, however, remains unchanged. Furthermore, for the MA scheme, $\Gamma=100$ achieves better system performance than $\Gamma=10$ due to the availability of more candidate MA positions, at the cost of increased computational complexity.

\begin{figure}[!t]
	\centering
	\includegraphics[width=1\linewidth]{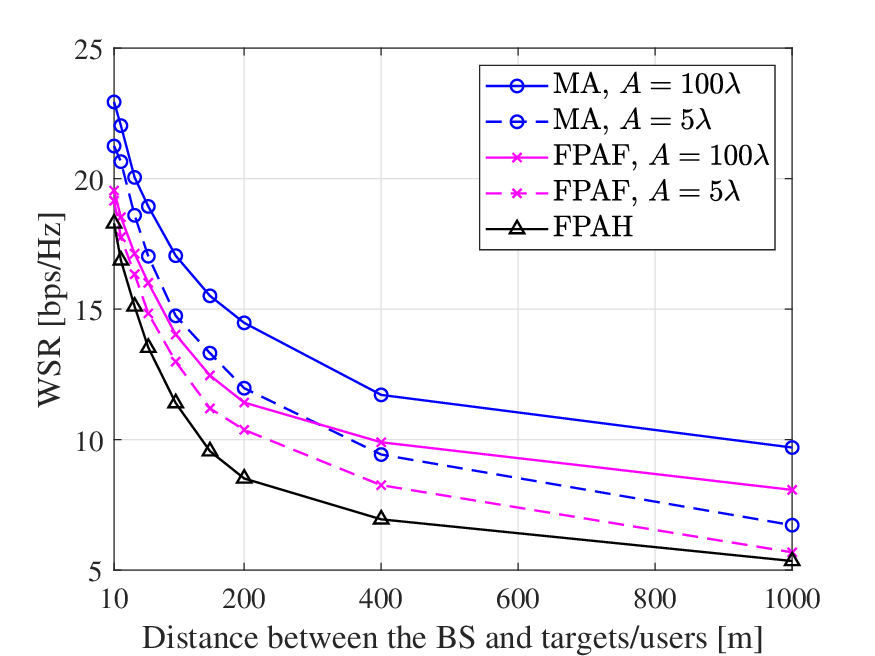}
	\caption{Comparison of WSR between the proposed and benchmark schemes w.r.t. the distance between the BS and targets/users.}
	\label{distance}
\end{figure}
In Fig. \ref{distance}, we compare the WSRs of the proposed and benchmark schemes w.r.t. the distance between the BS and targets/users. The distances between the BS and different targets/users are set within a 5-meter interval, centered around a specified distance. We can see that as the distance increases, the WSRs of all schemes decreases. This is because the increase in distance weakens system performance in the following two aspects: 1) increased path loss; and 2) the transition from near-field ISAC to far-field ISAC, which leads to the gradual loss of the distance dimension in the ISAC system. However, in the same communication scenario, the MA scheme still outperforms the FPAF and FPAH schemes due to the additional DoFs introduced by antenna position optimization.

\begin{figure}[!t]
	\centering
	\includegraphics[width=1\linewidth]{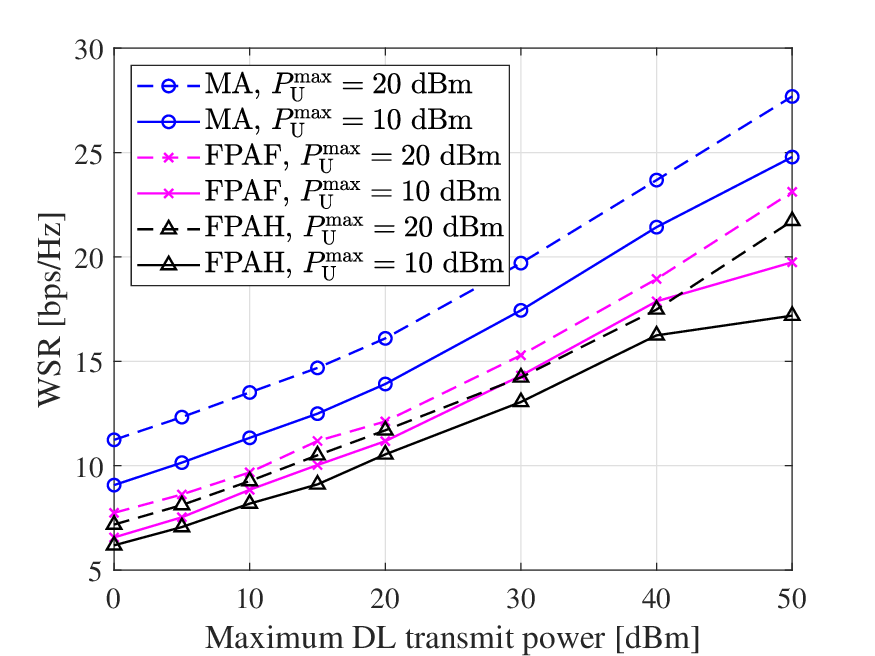}
	\caption{Comparison of WSR between the proposed and benchmark schemes w.r.t. the maximum DL transmit power.}
	\label{PD}
\end{figure}
Next, we show the WSRs of the proposed and benchmark schemes w.r.t. the maximum DL transmit power under different maximum UL transmit power constraints in Fig. \ref{PD}. Overall, the WSRs of all schemes improve as the maximum DL transmit power increases, since higher DL transmit power ensures more reliable DL transmission quality. However, the high DL transmit power can cause strong echo signals from the sensing targets and SI, which in turn degrade the SINRs of the UL users. Therefore, as shown in Fig. \ref{PD}, when $P^{\mathrm{max}}_{\mathrm{D}} \ge 40$ dBm, both the FPAF and FPAH schemes exhibit a noticeable slowdown in WSR increase compared to the MA scheme at low maximum UL transmit power case, i.e., $P^{\mathrm{max}}_{\mathrm{U}} = 10$ dBm. In addition, compared to FPA-based schemes, the MA scheme can save the DL transmit power at the same WSR level by leveraging flexible antenna movement.

\begin{figure}[!t]
	\centering
	\includegraphics[width=1\linewidth]{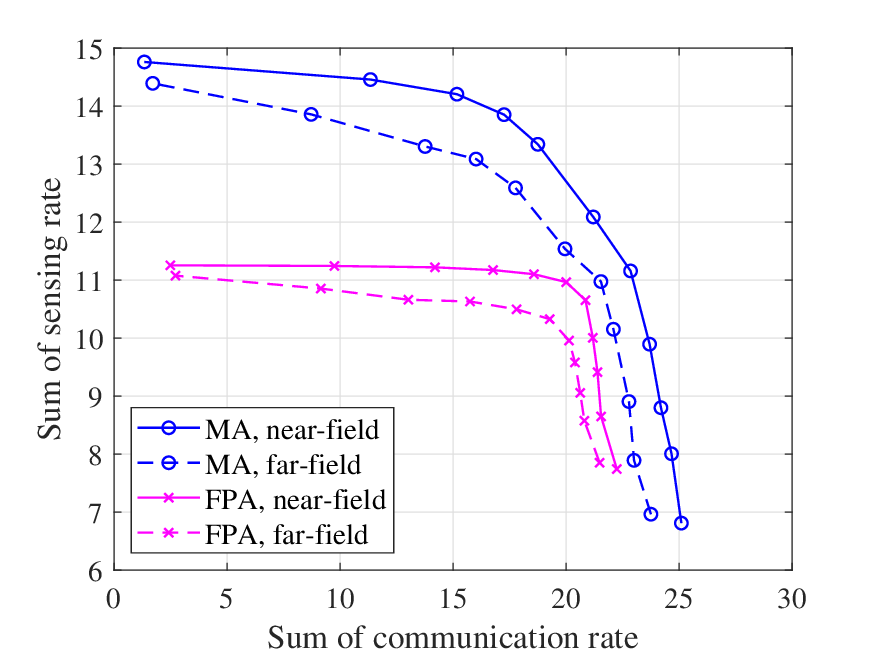}
	\caption{Trade-off between sensing and communication performance.}
	\label{trade_off}
\end{figure}
In Fig. \ref{trade_off}, we analyze the trade-off between sensing and communication. We set $P^\mathrm{max}_\mathrm{D}=30$ dBm, $\Gamma=10$, and ${\varpi _{{\mathrm{U}_j}}}={\varpi _{{\mathrm{D}_k}}}$. By adjusting the sensing rate weight ${\varpi _{{\mathrm{S}_l}}}$ and the communication weights ${\varpi _{{\mathrm{U}_j}}}$ and ${\varpi _{{\mathrm{D}_k}}}$, the trade-off curves between the sensing rate and the communication rate are plotted.  For example, a purely communication-oriented design can be realized by setting ${\varpi _{{\mathrm{S}_l}}}=0$. To illustrate the performance improvements enabled by the repositioning capability of MAs and the incorporation of near-field characteristics, we consider four typical ISAC system configurations: 1) MA, near-field: The moving region size is set to $A=100\lambda$, and thus the targets and users are located within the near-field region of the BS; 2) MA, far-field: The moving region size is set to $A=5\lambda$. In this case, the signal propagation between the targets/users and the BS follows the far-field model; 3) FPA, near-field: The antenna placement is determined based on the FPAF scheme with $A=100\lambda$, and thus the near-field condition consistent with that of 1) is ensured; and 4) FPA, far-field: The antenna placement is determined based on the FPAH scheme, under which the targets and users are located within the far-field region of the BS. As shown in Fig. \ref{trade_off}, the ISAC system achieves a better sensing-communication trade-off under near-field conditions compared to the far-field case. This improvement is attributed to the spherical-wave channel model, which enables the BS to distinguish targets and users at different distances and angles. Moreover, the MA system can exploit the additional DoF introduced by the antenna position optimization to flexibly reposition the antennas, thereby further enlarging the achievable sensing-communication trade-off region compared to the FPA system in both far-field and near-field scenarios.

\begin{figure}[!t]
	\centering
	\includegraphics[width=1\linewidth]{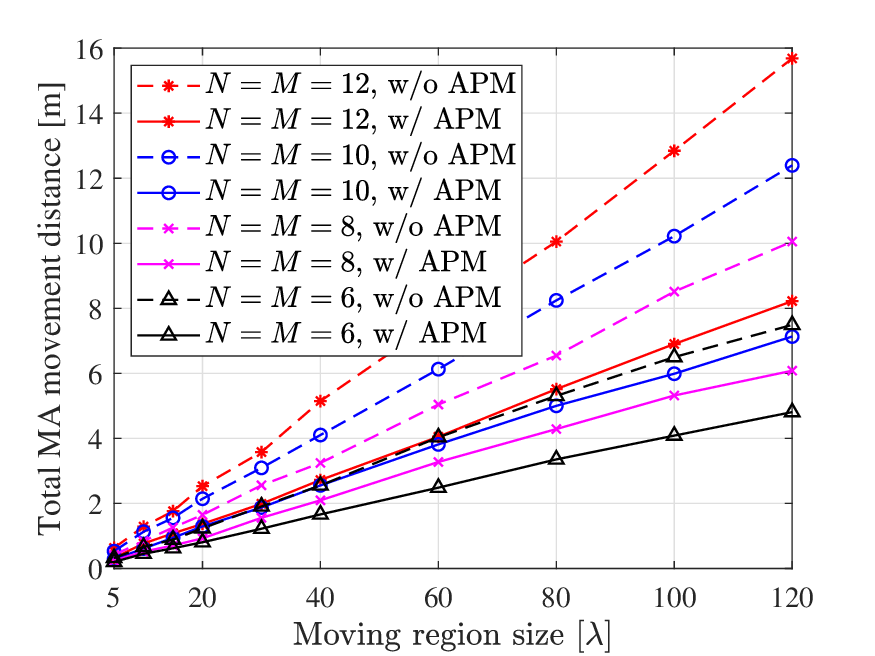}
	\caption{Total MA movement distance for schemes with (w/) or without (w/o) APM algorithm w.r.t. the moving region size.}
	\label{move_dis}
\end{figure}
Finally, Fig. \ref{move_dis} illustrates the total MA movement distance for schemes with (w/) or without (w/o) APM algorithm w.r.t. the moving region size under different numbers of MAs. It can be seen that the total MA movement distance increases linearly with the size of the moving region, and the APM algorithm significantly reduces this distance. Moreover, as the number of MAs increases, the gap between the schemes w/ and w/o APM algorithm widens. Specifically, when the moving region size is $120\lambda$, the proposed APM algorithm reduces the total MA movement distance by 35.79\%, 39.50\%, 42.46\%, and 47.59\% for $N=M=$ 6, 8, 10, and 12, respectively. Typically, in MA-aided near-field ISAC systems, the size of moving region and the number of MAs are large. Therefore, solely considering antenna position optimization without accounting for antenna's position matching with their initial positions leads to long MA movement distance, thereby increasing the time delay and hardware burden.
\section{Conclusion}\label{5}
In this paper, we investigated the MA-aided near-field ISAC system. We characterized the near-field sensing and communication channels w.r.t. MA positions using the spherical wave model and formulated a joint optimization problem to maximize the system's WSR achievable for both communications and sensing. To solve this non-convex optimization problem, we proposed a two-layer RP algorithm where multiple MA positions were randomly initialized. For each MA position, the beamformers, sensing signal covariance matrices, and UL power allocation were iteratively optimized using the AO algorithm until convergence. The MA position that achieves the maximum WSR was then selected as the optimized position. Moreover, considering the large size of moving region in near-field MA systems, we proposed an APM algorithm based on the greedy strategy to reduce the total MA movement distance, thereby alleviating the cost of antenna movement. Simulation results verified the effectiveness of the proposed algorithms and the advantages of the considered MA-aided ISAC scheme compared to conventional FPA-based schemes. Furthermore, the results showed that equipping the BS with large regions for free MA movement increases the equivalent array aperture, thereby significantly expanding the near-field region without requiring more antennas or higher frequencies. Compared to the far-field ISAC system, the additional distance dimension introduced by the near-field ISAC system can enhance system performance for multi-target sensing and multi-user communication through precise beamfocusing. This can be more efficiently exploited by optimally designing the MAs' positions matching the ISAC channels. 
\section*{Acknowledgments}
The calculations were supported by the High-Performance Computing Platform of Peking University.

\end{document}